\documentclass{article}

\usepackage{amsmath}  
\usepackage[english]{babel}
\usepackage[utf8]{inputenc}
\usepackage{amssymb}
\usepackage{graphicx}
\usepackage{authblk}
\usepackage{caption}
\usepackage{multirow}
\usepackage{amscd}
\usepackage{geometry}
\usepackage{a4wide}
\usepackage[usenames]{color}
\usepackage{etaremune} 
\usepackage[hidelinks]{hyperref}
\usepackage{amsfonts}
\usepackage{ifthen}
\usepackage{color}
\usepackage{amscd}
\usepackage{latexsym}
\usepackage{subfigure}
\usepackage{mathrsfs}
\usepackage{url}

\newcommand{\bd}{\begin{definition}}
	\newcommand{\ed}{\end{definition}}
\newcommand{\bt}{\begin{theorem}}
	\newcommand{\et}{\end{theorem}}
\newcommand{\bi}{\begin{itemize}}
	\newcommand{\ei}{\end{itemize}}
\newcommand{\ben}{\begin{enumerate}}
	\newcommand{\een}{\end{enumerate}}
\newcommand{\beq}{\begin{equation}}
	\newcommand{\eeq}{\end{equation}}

\newcommand{\R}{\mbox{$ \mathbb{R}  $}}

\newtheorem{definition}{Def.}[section]
\newtheorem{theorem}{Theorem}[section]
\newtheorem{proposition}{Proposition}[section]

\newcommand*\bell{\ensuremath{\boldsymbol\ell}}
\newcommand*\biota{\ensuremath{\boldsymbol\iota}}
\newcommand{\hr}{\mbox{$ \mathcal{H}(2,\mathbb R) $}}
\newcommand{\spinf}{\mbox{$ \mathbb R \oplus \R^2 $}}

\DeclareMathOperator{\Tr}{Tr}

\title{A quantum information-based refoundation of color perception concepts}

\author[1]{Michel Berthier\thanks{michel.berthier@univ-lr.fr}}
\author[2]{Nicoletta Prencipe\thanks{nicoletta.prencipe@math.u-bordeaux.fr, also working for Huawei Technologies France SASU}}
\author[2]{Edoardo Provenzi\thanks{edoardo.provenzi@math.u-bordeaux.fr, corresponding author}}
\affil[1]{Laboratoire MIA, Pôle Sciences et Technologie, Université de La Rochelle, 23 Avenue A. Einstein, BP 33060, 17031 La Rochelle Cedex, France}
\affil[2]{Université de Bordeaux, CNRS, Bordeaux INP, IMB, UMR 5251\\ F-33400, 351 Cours de la Libération, Talence, France}

\date{}

\begin{document}
	
	\maketitle
	
	
	\begin{abstract}  
	In this paper we deal with the problem of overcoming the intuitive definition of several color perception attributes by replacing them with novel mathematically rigorous ones. Our framework is a quantum-like color perception theory recently developed, which constitutes a radical change of view with respect to the classical CIE models and their color appearance counterparts. We show how quantum information concepts, as e.g. effects, generalized states, post-measurement transformations and relative entropy provide tools that seem to be perfectly fit to model color perception attributes as brightness, lightness, colorfulness, chroma, saturation and hue. An illustration of the efficiency of these novel definitions is provided by the rigorous derivation of the so-called lightness constancy phenomenon.
	\end{abstract}

\section{Introduction}\label{sec:intro}
The foundation of every scientific theory relies on the precise choice of the ideal model through which intuitive concepts are converted into mathematical entities, which must be \textit{rigorously defined} and whose properties are assumed to be the primitive axioms of the theory.  

As we shall discuss more thoroughly later, while the axioms of color perception theory are well-established, several definitions of color attributes still remain at an intuitive level: as it can be read in \cite{Wyszecky:82} or \cite{Fairchild:13}, paramount references in this domain, the classical color perception theory based on the CIE (Commission Interntional de l'\'Eclairage) construction is riddled by sloppy or circular definitions, as reported in section \ref{sec:vocabulary}. A quite recent example of misleading use of chromatic terms is the excerpt: `\textit{the chroma saturation level}', repeated 15 times in the paper  \cite{Kawashima:2017}, which appears in the \textit{official proceedings} of the 11th Biennial Joint meeting of the CIE/USA and CNC/CIE. 

It is evident that the lack of precision in these definitions poses not only a serious problem for the theoretical foundation of a color perception model, but also for practical applications and measurements. These considerations justify the interest in trying to setting up a mathematically rigorous vocabulary of perceptual color attributes, which is the main aim of the present paper. 

It is important to stress that we shall not provide this formalization within the classical CIE color spaces LMS, RGB, sRGB, XYZ, etc. or using their color appearance counterparts CIELuv, CIELab, CIECAM, and so on. The reason why we avoid considering the first class of color spaces is that they are well-known not to be fit for a perceptual analysis of color, see e.g. chapter 6 in \cite{Fairchild:13}, having been constructed under extremely constrained conditions and modeled using merely the cone photoreceptors sensitivities, without taking into account neither the fundamental role of chromatic opposition, nor the post-retinal brain functions. As a consequence, they totally fail to explain color appearance phenomena.

This failure was the motivation to built the so-called color appearance spaces, which, however, share the same unfit basis, being non-linear transformations of the XYZ space. Apart from the fact that it hardly makes sense to build perceptual attributes from non-perceptual ones, the non-linear functions used to transform the XYZ coordinates were determined empirically and ad-hoc parameters had to be introduced in order to fit experimental data obtained, again, in very restrictive controlled conditions. In \cite{Koenderink:2000}, Koenderink and van Doorn vividly describe the current state of the art on colorimetry as follows: `\textit{As the field is presented in the standard texts it is somewhat of a chamber of horrors: colorimetry proper is hardly distinguished from a large number of elaborations (involving the notion of `luminance' and of absolute color judgments for instance) and treatments are dominated by virtually ad hoc definitions (full of magical numbers and arbitrarily fitted functions). We know of no text where the essential structure is presented in a clean fashion. Perhaps the best textbook to obtain a notion of colorimetry is still Bouma's of the late 1940’s}'.

Sharing with Koenderink and van Doorn the exigence of mathematical rigor and the skepticism about the scientific basis on which modern colorimetry is founded, we advocate the need of a firm refoundation of color perception theory. We have found the source for our program in old works, not the excellent book  \cite{Bouma:71}, but the work of the great scientists that predated the CIE era, i.e. Newton, Maxwell, Young, von Helmoltz, Hering, Ostwald and Schr\"odinger. 

In fact, in 1920, Schr\"odinger identified what he thought to be the minimal set of axioms needed to fully describe the perception (by a normal trichromat) of a color in isolation from the rest of a visual scene. These axioms can be resumed by saying that the set $\cal C$ of perceptual colors is not a simple collection of sensations, but it has the structure of a 3-dimensional regular convex cone, see \cite{Schroedinger:20}. 

In 1974, Resnikoff noted that the class of color spaces identified by Schr\"odinger was too large to single out a well-defined geometry. Searching for another property to  complete Schr\"odinger's axioms, he exhibited a transitive group action on $\cal C$ that makes it a homogeneous space, see \cite{Resnikoff:74} or \cite{Provenzi:20} for more details. This result has the crucial consequence that only two geometric structures can be compatible with a 3-dimensional regular convex homogeneous cone: the first is trivial, i.e. $\R^+\times \R^+\times \R^+$, which is the abstract geometrical prototype of the LMS, RGB and XYZ spaces; the second is $\R^+\times \textbf{H}$, where $\textbf{H}$ is a 2-dimensional hyperbolic model with constant curvature $-1$.  Resnikoff also showed that these spaces agree with the so-called positive cones of the only two non-isomorphic formally real Jordan algebras of dimension 3, as we will briefly recall in section \ref{sec:recapquant}. 

The profound relationship between Jordan algebras and quantum theories, together with other motivations that will be discussed in section \ref{sec:recapquant}, led to the investigation of an alternative color perception theory in  \cite{BerthierProvenzi:19,Berthier:2020,Berthier:2021JofImaging,Berthier:21JMIV,Berthier:21JMP,Guennec:21,BerthierProvenzi:21GSI,BerthierProvenzi:2022PRS}. 

The result is a rigorous  mathematical theory free from incoherences, that permits to: reconcile in a natural way trichromacy with Hering's opponency \cite{Berthier:2020,Berthier:21JMIV}; formalize Newton's chromatic disk \cite{Berthier:2020}; 
single out the Hilbert-Klein hyperbolic metric as a natural perceptual chromatic distance \cite{Berthier:21JMP};  solve the long-lasting problem of bounding the infinite perceptual color cone to a convex finite-volume solid of perceived colors \cite{Berthier:2020,BerthierProvenzi:2022PRS}; predict the existence of uncertainty relations for chromatic opposition \cite{Berthier:2021JofImaging}; implement white balance algorithms by means of Lorentz boosts \cite{Guennec:21,BerthierProvenzi:2022PRS}.

In this paper, in order to provide a rigorous vocabulary of color perception attributes, we will exploit in particular the results obtained in \cite{BerthierProvenzi:2022PRS} thanks to the use of tools coming from quantum information theory, such as generalized quantum states, L\"uders transformations and effects.

The plan of the paper is the following: in section \ref{sec:vocabulary} we recall the classical CIE nomenclature for color attributes, underlying its intuitive status; section \ref{sec:recapquant} provides an essential recap of the quantum-like theory of color perception from emitting sources of light that supplies the setting for the novel results of the paper, by underlying, in particular, the fundamental role of quantum effects and the surprising usefulness of relativistic transformations induced by them in white balance algorithms; section \ref{sec:obsillpercolor} starts our novel contributions by defining the concept of observer, patch, illuminant and by identifying a perceived color with a post-measurement generalized state; in section \ref{sec:lightBright} we propose rigorous definitions of  brightness and lightness coherent with the quantum-like framework; in section \ref{sec:chromattributes} we deal with chromatic attributes as chroma, colorfulness, hue and saturation, stressing the vital role played by chromatic opposition and relative quantum entropy; section \ref{sec:lightconstacr} is an illustration of the potential use of our new system of definitions on the specific example of the phenomenon of lightness constancy; finally, in the conclusions we discuss the relevance of our results in image processing.

\section{Classical glossary of color perception attributes}\label{sec:vocabulary}

The following list provides the official definitions, that we quote \textit{verbatim}, of color perceptual attributes, see e.g. chapter 6 (page 487) of \cite{Wyszecky:82}, chapter 4 of \cite{Fairchild:13}, or the official website  \url{https://cie.co.at/e-ilv}.

\begin{itemize}
	\item \textit{Color}: is that aspect of visual perception by which an observer may distinguish differences between two structure-free fields of view of the same size and shape, such as may be caused by differences in the spectral composition of the radiant energy concerned in the observation.
	
	\item \textit{Related color}: it is a color perceived to belong to an area or object seen \textit{in relation to other colors}.
	
	\item \textit{Unrelated color}: it is a color perceived to belong to an area or object seen \textit{in isolation from other colors}.
	
	\item \textit{Hue}: is the attribute of a color perception denoted by blue, green, yellow, red, purple and so on. \textit{Unique hues} are hues that cannot be further described by the use of the hue names other than its own. There are four unique hues: red, green, yellow and blue. The \textit{hueness} of a color stimulus can be described as combinations of two unique hues; for example, orange is yellowish-red or reddish-yellow. Nonunique hues are also referred to as \textit{binary hues}.
	
	\item \textit{Chromatic color}: it is a color perceived possessing hue.
	
	\item \textit{Achromatic color}: it is a color perceived devoid of hue.
	
	\item \textit{Brightness}: attribute of a visual sensation according to which an area appears to be more or less intense; or, according to which the area in which the visual stimulus is present appears to emit more or less light. Variations in brightness range from \textit{bright} to \textit{dim}.
	
	\item \textit{Lightness}: attribute of a visual sensation according to which the area in which the visual stimulus is presented appears to emit more or less light in proportion to that emitted by a similarly illuminated area perceived as a white stimulus. In a sense, lightness may be referred to as \textit{relative brightness}. Variations in lightness range from \textit{light} to \textit{dark}. 
	
	\item \textit{Colorfulness}: attribute of a visual sensation according to which the perceived color of an area appears to be more or less chromatic.
	
	\item \textit{Chroma}: attribute of a visual sensation which permits a judgment to be made of the degree to which a chromatic stimulus differs from an achromatic stimulus of the same brightness. In a sense, chroma is \textit{relative colorfulness}.
	
	\item \textit{Saturation}: attribute of a visual sensation which permits a judgment to be made of the degree to which a chromatic stimulus differs from an achromatic stimulus regardless of their brightness.
	
\end{itemize}

The naive nature of the previous definitions is evident. In \cite{Fairchild:13}, the relationship between some of the attributes defined above is resumed in the following \textit{intuitive} equations:

\begin{equation}\label{eq:light}
	\text{Lightness}= \frac{\text{Brightness}}{\text{Brightness(White)}},
\end{equation}
where `White' refers of course to a surface that is perceived as white.

\begin{equation}\label{eq:Chroma}
	\text{Chroma}=\frac{\text{Colorfulness}}{\text{Brightness(White)}},
\end{equation}

\begin{equation}\label{eq:Saturation}
	\text{Saturation}=\frac{\text{Colorfulness}}{\text{Brightness}}.
\end{equation}

There exist analytical formulae to express attributes as hue, saturation, chroma and so on both in the classical CIE spaces and in the color appearance ones, see e.g. \cite{Wyszecky:82,Gonzalez:02}, which gave rise to a plethora of color spaces, as e.g.  HSL, HSV, HSI, LCh(ab), LCh(uv) and so on.

These definitions of color attributes, however, have been formulated at the expense of scientific rigor. We quote just two instances that should suffice to motivate the previous sentence: the first, a clear example of data-fitting with no theoretical model behind and plagued by several parameters, is given by the definition of the $L$ coordinate in the CIELab space: $L=116  f\left(Y/Y_n\right) - 16$, where 
\beq 
f(t) = \begin{cases}
	\sqrt[3]{t} & \text{if } t > \delta^3 \\
	\frac{t}{3 \delta^2} + \frac{4}{29} & \text{otherwise}
\end{cases}, \quad	\delta = \tfrac{6}{29},
\eeq 
and $Y_n$ is the $Y$ coordinate of an achromatic illuminant which serves as reference. The second instance is the definition of chroma in the CIELab space, which is taken to be the Euclidean norm of the chromatic vector with components $(a,b)$, i.e. $C_{ab} := \sqrt{a^{2} + b^{2}}$. This is a clear example of a definition motivated just by computational convenience at the expense of coherence with the many theoretical and experimental evidences about the hyperbolic nature of chromatic attributes, see e.g. \cite{Farup:14} or \cite{Berthier:21JMP} and the references therein. 

Having motivated the exigence of a scientifically precise treatment of color perception features and why we cannot rely on the CIE constructions, we now proceed with a recap of the quantum-like model that will provide the alternative setting of our analysis.

\section{An essential recap about a quantum-like theory of color perception}\label{sec:recapquant}

For the sake of self-consistency and linear development of the present paper, we recap in this section the most important results of the quantum-like theory of color perception that has been built in the papers quoted in the introduction, together with some novel contributions that will be duly specified. 

The starting point of our theory is the \textit{quantum trichromacy axiom} \cite{Berthier:2020}, which states that \textit{the space of perceptual unrelated colors sensed by a normal trichromatic observer is the domain of positivity of a non-associative 3-dimensional formally real Jordan algebra} (FRJA from now on). The classification of FRJAs guarantees that non-associative 3D FRJAs are all  isomorphic to either the so-called \textit{spin factor}\footnote{As a vector space, $\spinf$ can be identified with the 3-dimensional Minkowski space $\R^{1,2}$. $\spinf$ and $\hr$ become Jordan algebras when endowed with suitable non-associative bilinear products whose specification is not important in the present paper, we refer the reader to the references \cite{Baez:12,Faraut:94,Mccrimmon:1978} for more details about Jordan algebras.} $\spinf$ or to the algebra of symmetric $2\times 2$ real matrices $\hr$, which, moreover, happen to be naturally isomorphic as Jordan algebras via the following transformation:
\begin{equation}\label{eq:isoH2}
	\begin{array}{cccl}
		\chi: & \mathcal H(2,\mathbb R)  & \stackrel{\sim}{\longrightarrow} & \mathbb R\oplus \mathbb R^2  \\
		& \begin{pmatrix}
			\alpha + v_1 & v_2 \\
			v_2 & \alpha - v_1
		\end{pmatrix}  & \longmapsto & (\alpha,(v_1,v_2))^t=\begin{pmatrix}
			\alpha \\ {\bf v}
		\end{pmatrix},\quad \alpha\in \R, \; {\bf v}=(v_1,v_2)^t\in \R^2,
	\end{array}
\end{equation}
which induces the following isomorphism between their domains of positivity, i.e. the set of their squares:
\beq\label{eq:cones1}
\begin{array}{ccc}
	\overline{\mathcal C}(\mathbb R \oplus \mathbb R^2) & = & \overline{\mathcal L^+}\\
	\rotatebox{90}{$\cong$} &    & \rotatebox{90}{$\cong$} \\
	\overline{\mathcal C}(\mathcal H(2,\mathbb R)) & = & {\overline{\mathcal H^+}(2,\mathbb R)}   
\end{array},
\eeq 
where $\overline{\mathcal L^+}:=\{(\alpha,{\bf v})^t\in \R^{1,2}, \; \alpha\ge 0, \;\alpha^2 - \|\textbf{v}\|^2\ge 0 \}$ is the closure of the \textit{future lightcone}, $\| \; \|$ is the Euclidean norm, and $\overline{\mathcal H^+}(2,\mathbb R)$ is the set of positive semi-definite $2\times 2$ matrices, i.e. symmetric matrices with non-negative trace and determinant.

A crucial property of any FRJA $\mathcal A$ is that $\overline{\cal C}(\mathcal A)$ is always self-dual, i.e.
\beq\label{eq:dualcone}
\overline{\mathcal C}(\mathcal A)\cong \overline{\mathcal C}^*(\mathcal A):=\{ \omega \in \mathcal A^* \; : \; \forall b\in \overline{\mathcal C}(\mathcal A), \; \omega(b) \ge 0 \},
\eeq 
where $\mathcal A^*$ denotes the dual vector space of $\cal A$.

Non-associative Jordan algebras have been proven to provide a perfectly valid framework to develop quantum theories in the pioneering paper \cite{Jordan:34}, in the sense that their algebraic description of states and observables is equivalent to the density matrix formalism that can be constructed starting from the ordinary Hilbert space formulation, see e.g.  \cite{Townsend:85,Emch:2009}. Non-commutativity of Hermitian operators on a Hilbert space is replaced by non-associativity in the Jordan framework, this is essential to preserve the core of quantum theories, i.e. the existence of uncertainty relations, which cannot appear if the Jordan algebra of observables is both commutative and associative. 

Color perception shares at least two features with quantum theories: first, it makes no sense to talk about color in absolute terms, a color exists only when it is observed in well-specified observational conditions, see e.g. \cite{Wittgenstein:77,Russell:2001}; second, repeated color matching experiments on identically prepared visual scenes do not lead to a sharp selection of a color that matches the test, but to a distribution of selections picked around the most probable one, which is clearly reminiscent of the probabilistic interpretation of quantum mechanics. 

The quantum trichromacy axiom implies a radical change of paradigm with respect to classical colorimetry: we no more deal with color in terms of three coordinates belonging to a flat color space, but with a theory of \textit{color states and observables in duality with each other} in which, as we will point out soon, \textit{perceived colors are inextricably associated with measurements}, mathematically expressed by the so-called \textit{effects}. 

A perceptual observable of a visual scene, or simply an \textit{observable} $a$, is a sensation that can be measured leading to the registration of an outcome belonging to a certain set that depends on the observable. The algebra of observables $\mathcal A$ of our quantum-like theory of color perception is $\hr\cong \spinf$ and perceptual colors are particular observables that belong to their domain of positivity.

A perceptual state, or simply a \textit{state} $\bf s$, coincides, in practice, with the preparation  of a visual scene for the measurements of its observables. Two important examples of distinct states are the following: 
\begin{itemize}
	\item[a)] a \textit{color state from a light stimulus} is prepared by allowing a naturally or artificially emitted visible radiation to be perceived by an observer;
	\item[b)] a \textit{color state from an illuminated surface} is prepared by illuminating a colored patch so that it can be  perceived by an observer.
\end{itemize}
The color sensation measurement in both states can be performed either \textit{qualitatively}, as an incommunicable sensation in the observer's mind, or \textit{quantitatively}, via a color matching experiment which produces an outcome that can be registered. Colorimetry can exist thanks to the latter form of measurement, first scientifically formalized by J. Clerk Maxwell.

Of course, the match must be coherent with the nature of the color state, so, in the case a), the observer must match the sensation generated by the emitting source of light with another visible radiation. Instead, in the case b), the observer must find another colored patch perceptually indistinguishable from the first one when lit by the same illuminant.

It is important to stress that the quantitative color matching experiments that led to the formulation of Schrödinger's  axioms were done with emitting sources of light, hence the results of the model that we are going to recall in the following subsections are valid in such conditions. Starting from section \ref{sec:obsillpercolor} we will extend them also to the perception of surface colors.

\subsection{Chromatic states and related entropies}\label{subsec:chromentropies}

In the algebraic formulation of quantum mechanics states are described by \textit{density matrices}, i.e. unit-trace positive semi-definite matrices. In the quantum-like theory of color perception, the \textit{chromatic state vectors} $\textbf{v}_ {\bf s}=(s_1,s_2)^t$ belonging to the unit disk $\mathcal D$ parameterize each density matrix $\rho_ {\bf s}$, in fact the perceptual chromatic state space can be identified with:
\beq\label{eq:dmat}
\mathcal S(\mathcal H(2,\R)) = \left\{ \rho_ {\bf s}\equiv \frac{1}{2}\begin{pmatrix}
	1+s_1 & s_2 \\ s_2 & 1-s_1
\end{pmatrix}, \;  \|\textbf{v}_ {\bf s}\|\le 1 \right\},
\eeq 
or, as a consequence of (\ref{eq:isoH2}), 
\begin{equation}
	\mathcal S(\mathcal \spinf) :=\chi( \mathcal S(\mathcal H(2,\R)))= \left\{ \chi(\rho_ {\bf s})=\frac{1}{2}\begin{pmatrix}
		1 \\ {\bf v}_{\bf s}
	\end{pmatrix}, \; \Vert{\bf v}_{\bf s}\Vert\leq 1 \right\}. 
\end{equation}
This is the state space of a \textit{rebit}, the $\R$-version of a qubit, see e.g. \cite{Wootters:2014}, and it happens to be the easiest known quantum system.

In the rest of the paper, to simplify the notation, we will identify a state $\bf s$ with the unique associated density matrix $\rho_{{\bf s}}\in \hr$ and vector $\chi(\rho_{{\bf s}})\in \spinf$. 

The \textit{expectation value} of an observable $a\in \hr$ on the state $\bf s$ is the average outcome of repeated and independent measurements of $a$ performed when the system is identically prepared in the state $\bf s$. It is given by:
\beq\label{eq:expval}
\langle a \rangle_{\bf s}=\Tr(\rho_{\bf s} \, a).
\eeq 
Polar coordinates are the most natural ones in $\mathcal D$ and they provide this alternative parameterization of the generic density matrix:
\begin{equation}\label{eq:rhortheta}
	\rho_ {\bf s}(r,\vartheta)=\frac{1}{2}\begin{pmatrix}
		1+r\cos\vartheta&r\sin\vartheta\\r\sin\vartheta & 1-r\cos\vartheta \end{pmatrix}, \qquad r\in [0,1], \; \vartheta\in [0,2\pi).
\end{equation}
States can be either \textit{mixed} or \textit{pure}, accordingly to the fact that they can be written as a convex combination of other states or not, respectively.  The two most commonly used descriptors of mixedness of a quantum state are the so-called \textit{von Neumann and linear entropies}. The normalized\footnote{The non-normalized von Neumann entropy is defined by replacing $\log_2$ with $\log$, in that case the maximal value that it reaches is not 1 but $\log 2$.} von Neumann entropy of a mixed state ${\bf s}$ is defined by:
\beq 
S(\rho_ {\bf s}):=-\Tr(\rho_ {\bf s} \log_2 \rho_ {\bf s})=-\sum\limits_{k}\lambda_k \log_2 \lambda_k,
\eeq 
where the numbers $\lambda_k$ are the strictly positive eigenvalues of $\rho_ {\bf s}$, repeated as many times in the sum as their algebraic multiplicity. In the case of the density matrix $\rho_{{\bf s}}(r,\vartheta)$ they are $\lambda_1= (1 - r)/2$ and $\lambda_2=(1 + r)/2$.

{In this paper, as a novel contribution, alongside the von Neumann entropy we will consider also the normalized\footnote{The non-normalized linear entropy is defined without the presence of the factor 2 and the maximal value that it reaches is not 1 but $1/2$.} linear entropy $S_L$, which is a lower approximation of the von Neumann entropy $S$. It can be obtained by considering only the first (linear) term in the Taylor expansion of the matrix logarithm, leading to the following definition:
	\beq 
	S_L(\rho_ {\bf s}) :=2\left[1-\Tr(\rho_{\bf s}^2)\right]= 2\left(1-\sum\limits_{k}\lambda_k^2\right).
	\eeq 
	As proven in \cite{Heinosaari:2011} or \cite{Petz:2007}, both entropies are  invariant under orthogonal conjugation, which implies that they are \textit{radial functions}, moreover, they are \textit{concave} and, importantly, they provide the same characterization of pure states and of the maximally mixed state, denoted with $\rho_{\bf 0}$:
	\begin{itemize}
		\item $\rho_ {\bf s}$ is a pure state if and only if $S(\rho_ {\bf s})=S_L(\rho_ {\bf s})=0$;
		\item $\rho_{\bf 0}=\text{argmax}_{\rho_ {\bf s}} \, S(\rho_ {\bf s})=\text{argmax}_{\rho_ {\bf s}}\, S_L(\rho_ {\bf s})$.
	\end{itemize}
	Finally, both entropies induce partial orders on the state space which, however, are different. 
	
	A straightforward calculation shows that the von Neumann and linear entropies are expressed by:
	\beq
	S(r)=\begin{cases}
		-\frac{1-r}{2}\log_2\left(\frac{1-r}{2}\right)-\frac{1+r}{2}\log_2\left(\frac{1+r}{2}\right)& r\in [0,1)\\
		0 & r=1
	\end{cases},
	\eeq 
	and $S_L(r) =1-r^2$, which show that $\rho_{\bf0}=Id_2/2$, where $Id_2$ is the $2\times 2$ identity matrix, or equivalently, $\chi(\rho_{\bf0})=\frac{1}{2}(1,{\bf0})^t$, which means that the maximally mixed state is parameterized by the null vector, the center of $\cal D$, where $S(0)=S_L(0)=1$. 
	
	Since the highest degree of entropy is equivalent to the minimal amount of chromatic information, $\rho_{\bf0}$ is identified with the \textit{achromatic state}, denoted with $\bf{s_a}$.
	Instead, pure states are parameterized by the points of the border of $\mathcal D$ and are identified with the \textit{hues} of perceived colors:
	\beq
	\begin{split}
		\mathcal {PS}(\mathcal H(2,\R)) & = \left\{ \rho_ {\bf s}= \frac{1}{2}\begin{pmatrix}
			1+s_1 & s_2 \\ s_2 & 1-s_1
		\end{pmatrix}, \;  \|\textbf{v}_ {\bf s}\|= 1 \right\} \\
		& =\left\{ \rho_ {\bf s}= \frac{1}{2}\begin{pmatrix}
			1+\cos \vartheta & \sin \vartheta \\ \sin \vartheta & 1-\cos \vartheta
		\end{pmatrix}, \;  \vartheta\in [0,2\pi) \right\},
	\end{split}
	\eeq 
	or, equivalently,
	\begin{equation}
		\mathcal {PS}(\mathcal \spinf) :=\chi( \mathcal {PS}(\mathcal H(2,\R)))= \left\{ \chi(\rho_ {\bf s})=\frac{1}{2}\begin{pmatrix}
			1 \\ {\bf v}_{\bf s}
		\end{pmatrix}, \; \Vert{\bf v}_{\bf s}\Vert = 1 \right\}. 
	\end{equation}
	In section \ref{subsec:huerelentropy} we will give a further,  and more precise, argument in favor of the interpretation of pure states as hues.
	
	Recalling the intuitive definition of saturation quoted in section \ref{sec:vocabulary}, it is quite natural to define the saturation $\Sigma$ of the chromatic state $\rho_{\bf s}(r,\vartheta)$ as follows \cite{Berthier:2021JofImaging,BerthierProvenzi:2022PRS}:
	\begin{itemize}
		\item if one considers the normalized von Neumann entropy, then, for $r\in [0,1)$:
		\beq 
		\begin{split}
			S(r)& =-\frac{1-r}{2}\log_2\left(\frac{1-r}{2}\right)-\frac{1+r}{2}\log_2\left(\frac{1+r}{2}\right) \\
			& = 1- \left[\frac{1}{2}\log_2\left(1-r\right)-\frac{r}{2}\log_2\left(1-r\right)+\frac{1}{2}\log_2\left(1+r\right)+\frac{r}{2}\log_2\left(1+r\right)\right] \\
			& = 1- \left[\frac{1}{2}\log_2(1-r^2)+\frac{r}{2}\log_2\left(\frac{1+r}{1-r}\right)\right],
		\end{split}
		\eeq
		so
		\beq\label{eq:satvonneumann}
		\Sigma(r):=1-S(r) =\frac{1}{2}\log_2(1-r^2)+\frac{r}{2}\log_2\left(\frac{1+r}{1-r}\right), \quad r\in [0,1),
		\eeq
		and $\Sigma(1):=1$; 
		\item if one considers the normalized linear entropy, then:
		\beq 
		\Sigma_L(r):=1-S_L(r) =r^2,
		\eeq 
	\end{itemize}
	in this way we have $\Sigma(0)=\Sigma_L(0)=0$ and $\Sigma(1)=\Sigma_L(1)=1$.
	The plot of these two proposals for saturation is displayed in Fig. \ref{fig:twosat}, where it can be seen that $\Sigma_L(r)$ grows slightly faster than $\Sigma(r)$. Up to now, both definitions of saturation seem equally plausible, however, in section \ref{subsec:chromrelentropy} we will provide a strong argument in favor of $\Sigma(r)$, thus privileging the von Neumann entropy over its linearized versions to be used for the definition of color saturation.
	
	\begin{figure}[!ht]
		\centering
		\includegraphics[scale=0.2]{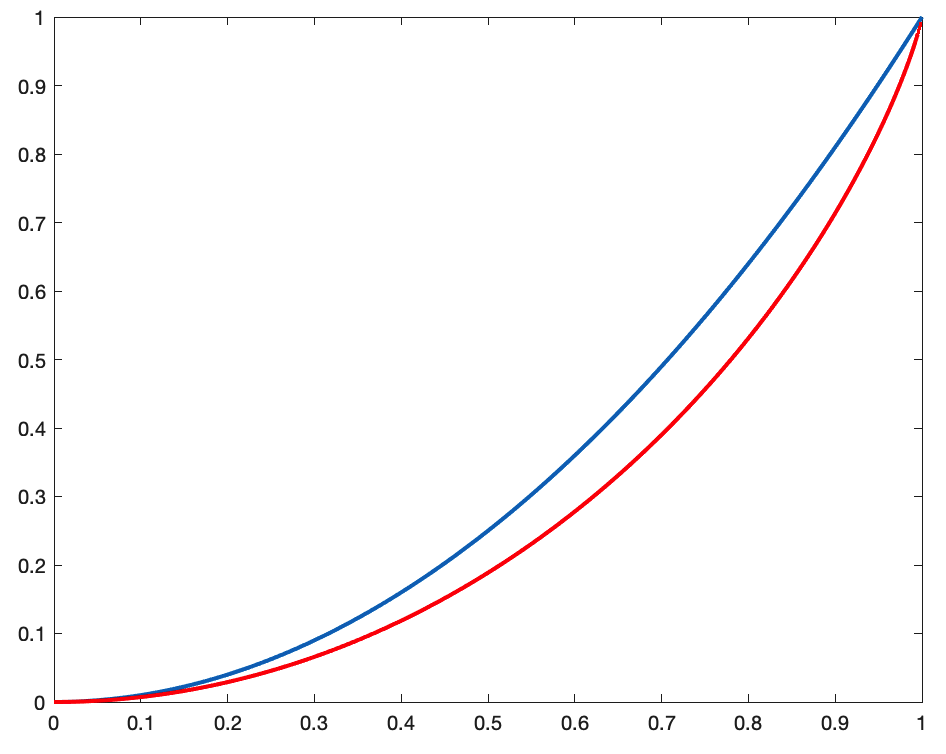}
		\caption{Graph of $\Sigma_L(r)$ (in blue) and of $\Sigma(r)$ (in red).}\label{fig:twosat}
	\end{figure}
	
	\subsection{Chromatic opponency: Hering's rebit}
	
	The next fundamental information to recall is how Hering's chromatic opponency naturally appears in the quantum-like formalism. The following presentation is novel with respect to our previous contributions and it is explicitly based on the canonical decomposition of density matrices known as \textit{Bloch representation}. Given $(e_i)_{i=0}^2$, the canonical basis of $\R\oplus \R^2$, if we define $\sigma_i:=\chi^{-1}(e_i)$, then we get 
	\begin{equation}
		\sigma_0\equiv Id_2, \ \ \sigma_1=\begin{pmatrix} 1 & 0 \\0 & -1\end{pmatrix},\ \ \sigma_2=\begin{pmatrix} 0& 1 \\1 & 0\end{pmatrix},
	\end{equation}
	where $\sigma_1$ and $\sigma_2$ can be recognized to be the two \textit{real Pauli matrices}. The generic density matrix of $\mathcal S(\mathcal H(2,\R))$ can be decomposed in terms of the real Pauli matrices as follows:
	\beq\label{eq:normaltri}
	\rho_{\bf s} (s_1,s_2)=\rho_{\bf 0} +{1\over 2}(s_1\sigma_1 + s_2\sigma_2)=\rho_{\bf 0} +{1\over 2} {\bf v}_{\bf s} \cdot \vec{\sigma},
	\eeq 
	where ${\bf v_s}=(s_1,s_2)^t$ is called the \textit{Bloch vector} associated to ${\bf s}$ and ${\bf v_s}\cdot \vec \sigma := s_1 \sigma_1 + s_2 \sigma_2$. The set $\{\sigma_0,\sigma_1,\sigma_2\}$ is an orthogonal basis for $\mathcal H(2,\R)$ with respect to the Hilbert-Schmidt inner product, i.e.
	\beq  
	\sigma_i \cdot \sigma_j := \Tr(\sigma_i\sigma_j)=2\delta_{ij}, \qquad i,j=0,1,2,
	\eeq 
	this allows us to identify the components of the Bloch vector with the expectation values of the real Pauli matrices on the state $\bf s$, in fact:
	\beq 
	{\bf v_s}=(s_1,s_2)=(\Tr(\rho_{\bf s} \, \sigma_1),\Tr(\rho_{\bf s} \, \sigma_2))=(\langle \sigma_1\rangle_{\bf s},\langle \sigma_2 \rangle_{\bf s}).
	\eeq  
	As a consequence, eq. \eqref{eq:normaltri} can be re-written as follows:
	\beq\label{eq:rhopauli}
	\rho_{\bf s} =\rho_{\bf 0} + \frac{1}{2}\begin{pmatrix}
		\langle \sigma_1\rangle_{\bf s} & \langle \sigma_2\rangle_{\bf s} \, \\  \langle \sigma_2\rangle_{\bf s} & \; - \langle \sigma_1\rangle_{\bf s}
	\end{pmatrix},
	\eeq 
	and its polar expression is:
	\beq\label{eq:rhopolar} 
	\rho_{\bf s}(r,\vartheta)=  \rho_{\bf 0} +\frac{1}{2} \left[r\cos \vartheta \, \sigma_1 + r\sin \vartheta \, \sigma_2\right],
	\eeq 
	with $r\in [0,1]$ and $\vartheta \in [0,2\pi)$. 
	
	Given two generic angles $\vartheta_1,\vartheta_1\in [0,2\pi)$, the pure states $\rho_{{\bf s}_k}(1,\vartheta_k)$, $k=1,2$, are rank-1 projectors that can be represented as follows:
	\beq \rho_{{\bf s}_k}(1,\vartheta_k)= {1\over 2}(Id_2+\cos \vartheta_k\sigma_1 + \sin \vartheta_k\sigma_2)
	\equiv \rho_{\bf 0}+{1\over 2}{\bf v}_{{\bf s}_k} \cdot \vec{\sigma}, \eeq 
	with ${\bf v}_{{\bf s}_k}=(\cos \vartheta_k,\sin \vartheta_k)^t$, $k=1,2$.
	
	In quantum theories, orthogonality is used to measure \textit{incompatibility between states}, and $\rho_{{\bf s}_1}(1,\vartheta_1)$, $\rho_{{\bf s}_2}(1,\vartheta_2)$ project on two orthogonal rays in $\R^2$ if and only if their Bloch vectors ${\bf v}_{{\bf s}_1},{\bf v}_{{\bf s}_2}$ are antipodal, see e.g. \cite{Heinosaari:2011}.
	
	In Hering's theory of color perception, see \cite{Hering:1878}, incompatibility between color sensations is called \textit{opposition}, for this reason two pure states $\rho_{{\bf s}_1}(1,\vartheta_1)$ and $\rho_{{\bf s}_2}(1,\vartheta_2)$ are said to be \textit{chromatically opponent} when $|\vartheta_1-\vartheta_2|=\pi$. The concept of opposition will play a fundamental role in section \ref{sec:chromattributes}.
	
	Let us immediately use opponency to corroborate our interpretation of $\rho_{\bf 0}$ as the achromatic state: it is easy to prove that the following formula holds
	\begin{equation}
		\rho_{\bf 0} ={1\over 4}\rho_{\bf s}(1,0)+{1\over 4}\rho_{\bf s}(1,\pi)+{1\over 4}\rho_{\bf s}\left(1,{\pi\over 2}\right)+{1\over 4}\rho_{\bf s}\left(1,{3\pi\over 2}\right),
	\end{equation}
	this shows that $\rho_{\bf 0}$ is the mixed state obtained as a convex combination, with exactly the same coefficients, of the balance between two  couples of pure opponent chromatic states. 
	
	Notice also that the real Pauli matrices can be expressed as follows:
	\beq\label{eq:sigmas}
	\sigma_1=\rho_{\bf s}(1,0)-\rho_{\bf s}(1,\pi), \quad 
	\sigma_2=\rho_{\bf s}\left(1,{\pi\over 2}\right)-\rho_{\bf s}\left(1,{3\pi\over 2}\right),
	\eeq 
	thus eq. \eqref{eq:rhopolar} implies the crucial formula
	\begin{equation}\label{eq:Hering3d}
		\rho_{\bf s}(r,\vartheta) = \rho_{\bf 0}+\frac{1}{2} \left\{r\cos\vartheta \left[\rho_{\bf s}(1,0)-\rho_{\bf s}(1,\pi)\right]+r\sin\vartheta \left[\rho_{\bf s}\left(1,{\pi\over 2}\right)-\rho_{\bf s}\left(1,{3\pi\over 2}\right)\right] \right\}.
	\end{equation}
	Eq. \eqref{eq:Hering3d} is the exact quantum analogue of Hering's representation of color sensations: the generic chromatic state ${\bf s}$ identified by the density matrix $\rho_{\bf s}(r,\vartheta)$ can be interpreted as the \textit{contribution of the achromatic state} $\rho_{\bf 0}$ \textit{and the balance between two couples of opponent chromatic states}, encoded by the real Pauli matrices $\sigma_1,\sigma_2$. A fundamental remark on color perception made by Hering is that, for unrelated colors, while \textit{the chromatic information is intrinsic}, \textit{the achromatic part can be determined only by means of comparisons with other colors}, see e.g. \cite{Hubel:95}. An observer can measure the degree of opposition red vs. green and yellow vs. blue of an unrelated color, see e.g. \cite{Jameson:55}, but, due to adaptation mechanisms of the human visual system, he or she cannot establish how bright or dim a perceived unrelated color is (apart from extreme situations close to the visible threshold or the glare limit, that we do not consider here). This ambiguity is represented in eq. \eqref{eq:Hering3d} by the fact that $\rho_{\bf 0}$ appears as a sort of `offset state', independent of the state $\bf s$.
	
	For clear reasons, we call \textit{Hering's rebit} the quantum-like system that we have just described. This latter can be thought as a mathematical formalization of Newton's chromatic disk, as depicted in Fig.  \ref{fig:NewtonHering}.
	
	\begin{figure}[!ht]
		\centering
		\includegraphics[scale=0.1]{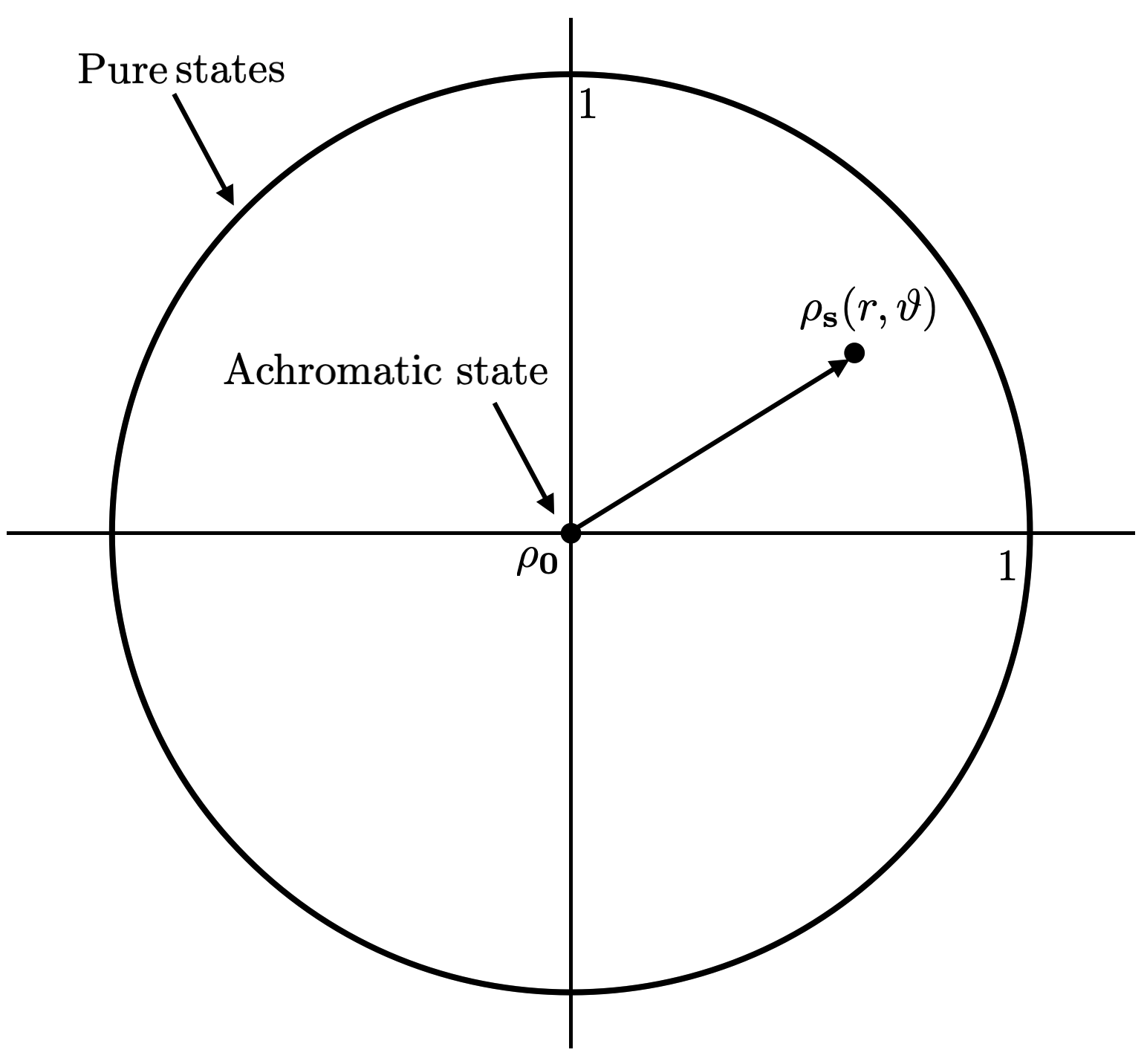}
		\caption{Hering's rebit as a quantum-like formalization of Newton's chromatic disk.}\label{fig:NewtonHering}
	\end{figure}
	
	To underline even more the fundamental role of opposition in creating the sensation of color, we end this section by quoting this sentence taken from  \cite{Devalois:1997}: `\textit{It is very misleading to consider the cones as color receptors or give them color names $[\dots]$ The specific information about color comes from lateral neural interactions which in every case involve a comparison of activity in different receptors $[\dots]$  although this fact is hidden by tenacious old theories and the continued use of color names for receptor types}'.
	
	\subsection{The fundamental role of quantum effects in modeling color measurement}
	
	The concept of \textit{effect} encodes the probabilistic nature of quantum measurements and lies at the very core of modern quantum theories, see e.g. the classical books  \cite{Kraus:83,Busch:97,Heinosaari:2011} for an overview on this fundamental topic.
	
	In sections \ref{subsubsec:colorsolid} and \ref{subsubsec:Luders} we will briefly recall how the concept of effect has been used to describe color measurement from emitting sources of light in previous papers. In section \ref{sec:obsillpercolor} we will extend these results to encompass also color measurements from an  illuminated surface.
	
	
	\subsubsection{Effects and the color solid of a real trichromatic  observer}\label{subsubsec:colorsolid}
	
	The quantum trichromacy axiom refers to an \textit{ideal} normal trichromatic observer, capable of a non-trivial response to light stimuli of any intensity, no matter how dim or intense. However, the visible threshold and glare limits, see e.g.  \cite{Koenderink:03,Provenzi:16modernphysics,Provenzi:20}, imply that the space of perceived colors perceived by a \textit{real} normal trichromatic observer is actually a finite convex subset, usually called \textit{color solid}, of the infinite cone $\overline{\mathcal C}(\mathcal A)$, $\mathcal A=\mathcal H(2,\R)$ or $\R \oplus \R^2$.
	
	As first argued in \cite{Berthier:2020}, a finite-volume color solid can be obtained in a natural way in the quantum-like framework by first re-writing $\overline{\mathcal C}(\mathcal A)$ to make states appear explicitly as follows
	\beq\label{eq:twoalpharho}
	\overline{\mathcal C}(\mathcal H(2,\R)) = \left\{2\alpha \rho_{\bf s}=\begin{pmatrix}
		\alpha (1+s_1) & \alpha s_2 \\ \alpha s_2 & \alpha (1-s_1)
	\end{pmatrix},\; \alpha\geq 0, \; {\bf v}_{\bf s}=(s_1,s_2)^t\in\mathcal D\right\} 
	\eeq
	and
	\beq\label{eq:twoalphasv}
	\overline{\mathcal C}(\R\oplus \R^2) = \left\{2\alpha\chi(\rho_{\bf s})=\begin{pmatrix}
		\alpha\\  \alpha {\bf v_s}
	\end{pmatrix},\; \alpha\geq 0, \; {\bf v_s}\in\mathcal D\right\},
	\eeq 
	and then by identifying an observed color with an \textit{effect}, which is defined to be an element $\eta_{\bf e}$ of $\overline{\mathcal C}(\mathcal H(2,\R))\cong  \overline{\mathcal H^+}(2,\R)$ bounded between the null and the identity $2\times 2$ matrix (with respect to the ordering of positive semi-definite matrices) or, equivalently, $\chi(\eta_{\bf e})\in \overline{\mathcal C}(\R\oplus \R^2)\cong \overline{\mathcal L^+}$. 
	
	It is useful to adopt a general symbol to denote an effect $\bf e$ when it is not important to know if it is realized as the matrix $\eta_{\bf e}$ or the vector $\chi(\eta_{\bf e})$. We will use the following notation:
	\beq
	{\bf e}:=(e_0,{\bf v_e}),
	\eeq 
	where $e_0$ and $\bf v_e$, called \textit{effect magnitude} and \textit{effect vector} play the role of $\alpha$ and $\bf v_s$ in eq. \eqref{eq:twoalpharho}, respectively. It is convenient to define the effect vector as follows:
	\beq
	{\bf v}_{\bf e}:=\left(\frac{e_1}{e_0},\frac{e_2}{e_0}\right)^t,
	\eeq 
	$e_0,e_1,e_2\in \R$, because then the matrix $\eta_{\bf e}$ can be written in this way 
	\begin{equation}\label{eq:mateff}
		\eta_{\bf e}=\left(\begin{array}{cc}e_0+e_1 & e_2 \\e_2 & e_0-e_1\end{array}\right),
	\end{equation}
	and  
	\begin{equation}\label{eq:chietae}
		\chi(\eta_{\bf e}):=e_0 \begin{pmatrix}
			1 \\ \textbf{v}_ {\bf e}
		\end{pmatrix}.
	\end{equation}
	
	
	The matrix $\eta_{\bf e}$ defines an effect if and only if ${\bf 0}\le \eta_{\bf e}\le \sigma_0$, this double inequality is equivalent to the request that the determinant and the trace of both $\eta_{\bf e}$ and $\sigma_0-\eta_{\bf e}$ are non-negative. From $\det(\eta_{\bf e})\ge 0$ we obtain ${\bf v}_{\bf e}\in \mathcal{D}$ and, by considering all the other constraints, we find that the effect space, or perceived color space, can be geometrically characterized in an explicit way as follows:
	\begin{equation}\label{eq:effect1}
		\mathcal E=\left\{(e_0,e_1,e_2)\in \R^3, \; e_0 \in [0,1], \;  e_1^2+e_2^2\le \min\limits_{e_0 \in [0,1]} \left\{(1-e_0)^2,e_0^2\right\} \right\}.
	\end{equation}
	$\mathcal E$ is a \textit{closed convex double cone} with a circular basis of radius $1/2$ located height $e_0=1/2$ and vertices in $(0,0,0)$ and $(1,0,0)$, associated to \textit{the null and the unit effect}, respectively \cite{BerthierProvenzi:2022PRS}. The geometry of $\mathcal E$ happens to be in perfect agreement with that of the perceived color spaces advocated by Ostwald and De Valois, see e.g. \cite{Devalois:2000}.
	
	
	The self duality of $\overline{\mathcal C}(\mathcal A)$ allows us to alternatively interpret effects as affine maps (that we indicate with the same symbol for simplicity) acting on chromatic states, see \cite{BerthierProvenzi:2022PRS} for more details:
	\begin{equation}\label{eq:effect2}
		\mathcal E\cong \left\{{\bf e}:{\cal S}(\mathcal A)\to[0,1], \; {\bf e}({\bf s})=e_0+e_1s_1+e_2s_2\right\},
	\end{equation} 
	${\bf e}({\bf s})$ is interpreted as the probability to register the outcome $(e_0,e_1,e_2)$ after a color measurement on the visual scene prepared in the state $\bf s$, i.e. ${\bf e}({\bf s})$ coincides with the expectation value $\langle {\bf e} \rangle_{\bf s}$, which can be written as follows:
	\begin{equation}\label{eq:expes} 
		\langle {\bf e} \rangle_{\bf s}={\rm Tr}(\rho_{\bf s} \, \eta_{\bf e})=e_0+e_1s_1+e_2s_2=e_0(1+{\bf v}_{\bf e}\cdot {\bf v}_{\bf s})=2\chi(\rho_{\bf s})\cdot \chi(\eta_{\bf e}).
	\end{equation}
	The so-called \textit{achromatic effect} is ${\bf e_a}:=e_0(1,{\bf 0})$, with  $e_0\in [0,1]$, it is characterized by a null effect vector ${\bf v_{e_a}}=\bf 0$, so that 
	\beq 
	\eta_{\bf e_a}=e_0 \sigma_0.
	\eeq 
	
	
	\subsubsection{Post-measurement generalized states}\label{subsubsec:Luders} 
	
	Effects parameterize a fundamental class of state transformations called \textit{Lüders operations}, which are \textit{convex-linear positive functions} $\psi_{\bf e}$ defined on the state space $\mathcal S(\hr)$ and satisfying the constraint: 
	\beq
	0\le \Tr(\psi_{\bf e}(\rho_{\bf s}))\le 1, \quad \text{for all } \rho_{\bf s}\in \mathcal S(\hr).
	\eeq  
	This implies that  $\mathcal S(\hr) \subset \psi_{\bf e}(\mathcal S(\hr)=:\tilde{\mathcal S}(\hr)$, i.e. $\rho_ {\bf s}$ will lose the property of having unit trace after a Lüders operation, becoming a so-called \textit{generalized density matrix} representing a \textit{post-measurement generalized state}. 
	From the identification between states and density matrices it follows that
	\beq\label{eq:gendensmat}
	\psi_ {\bf e}( {\bf s})\equiv \psi_{\bf e}(\rho_{\bf s})\in \tilde{\mathcal S}(\hr),
	\eeq  
	so
	\beq\label{eq:constraint01}
	\Tr(\psi_ {\bf e}(  {\bf s})) \in [0,1].
	\eeq 
	The analytical expression of the post-measurement generalized state $\psi_ {\bf e}(  {\bf s})$, see e.g. \cite{Busch:97} page 37, is:
	\beq\label{eq:psies}
	\psi_ {\bf e}(  {\bf s})=\eta_ {\bf e}^{1/2} \rho_ {\bf s} \eta_ {\bf e}^{1/2},
	\eeq 
	$\eta_ {\bf e}^{1/2}$ is called \textit{Kraus operator} associated to ${\bf e}$ and it is the square root of $\eta_ {\bf e}$, i.e. the only symmetric and positive semi-definite matrix such that $\eta_ {\bf e}^{1/2}\eta_ {\bf e}^{1/2}=\eta_ {\bf e}$. Thanks to the cyclic property of the trace,
	\beq\label{eq:expeff}
	\Tr(\psi_ {\bf e}(  {\bf s}))= \Tr(\rho_{\bf s} \, \eta_ {\bf e}) = \langle   {\bf e} \rangle_ {\bf s}=e_0(1+{\bf v}_{\bf e}\cdot {\bf v}_{\bf s}),
	\eeq
	so 
	\beq\label{eq:varphi}   
	\varphi_ {\bf e}(  {\bf s}):=\frac{\psi_ {\bf e}(  {\bf s})}{\langle   {\bf e} \rangle_ {\bf s}}
	\eeq  
	is a density matrix corresponding to a genuine state belonging to $\mathcal S(\hr)$.
	
	By convex-linearity, Lüders operations can be naturally extended to generalized states as follows:
	\beq\label{eq:psigen}
	\psi_{\bf e}(s_0 {\bf s})= s_0 \psi_{\bf e}({\bf s}), \qquad \forall s_0\in [0,1],
	\eeq 
	this implies 
	\beq\label{eq:oneplus} 
	\langle{\bf e}\rangle_{s_0\bf s}=\Tr(s_0\rho_{\bf s}\eta_{\bf e})=s_0\Tr(\rho_{\bf s}\eta_{\bf e}) = s_0 \langle {\bf e} \rangle_{\bf s} = e_0 s_0 (1+{\bf v_e} \cdot \bf{v_s}),
	\eeq 
	so
	\beq\label{eq:intrchr}
	\varphi_{\bf e}(s_0{\bf s})=\frac{\psi_{\bf e}(s_0{\bf s})}{\langle {\bf e}\rangle_{s_0\bf s}} = \frac{s_0\psi_{\bf e}({\bf s})}{s_0\langle {\bf e}\rangle_{\bf s}} = \varphi_{\bf e}({\bf s}),
	\eeq 
	thus the post-measurement chromatic state depends solely on $\bf s$ and not on $s_0$. This implies a formula that will be used often in the paper:
	\beq\label{eq:percareasplit}
	\psi_{\bf e}(s_0{\bf s}) =e_0 s_0 (1+{\bf v_e} \cdot \bf{v_s}) \, \varphi_{\bf e}({\bf s}).
	\eeq 
	This formula shows explicitly how the chromatic information about the state $\bf s$ and the expectation value of the effect $\bf e$ on $\bf s$ are \textit{fused together} in the post-measurement generalized state $\psi_{\bf e}(s_0{\bf s})$.
	
	In the case of an achromatic effect $\bf e_a$, for which ${\bf v_{e_a}}=\bf 0$, the previous formula gives
	\beq\label{eq:psiphiachr}
	\psi_{\bf e_a}(s_0{\bf s}) =e_0 s_0 \, \varphi_{\bf e_a}({\bf s}),
	\eeq 
	but $\eta_{\bf e_a}^{1/2}=\sqrt{e_0} \sigma_0$ so, by eq. \eqref{eq:psies}, 
	\beq\label{eq:psiachr}
	\psi_ {\bf e_a}(s_0{\bf s})=e_0s_0\rho_{\bf s},
	\eeq 
	hence $\varphi_{\bf e_a}({\bf s}) = \rho_{\bf s}$, or, by identifying $\rho_{\bf s}$ with the chromatic state $\bf s$,
	\beq\label{eq:achreffectstate} 
	\varphi_{\bf e_a}({\bf s}) = \bf s,
	\eeq 
	this means that the post-measurement state induced by the action of an achromatic effect coincides with the original state. 
	
	Remarkably, in \cite{BerthierProvenzi:2022PRS}, it has been shown that \textit{the state change $\bf s\mapsto \psi_{{\bf e}}({\bf s})$ induced by the act of observing a color} is  implemented through a 3-dimensional normalized Lorentz boost in the direction of $\bf v_e$. 
	
	As also proven in \cite{BerthierProvenzi:2022PRS}, the post-measurement chromatic state vector is the Einstein-Poincaré relativistic sum of $\bf v_e$ and $\bf v_s$, i.e. 
	\begin{equation}\label{eq:vectLuders}
		{\bf v}_{\varphi_{\bf e}({\bf s})} =   {\bf v_e} \oplus {\bf v_s},
	\end{equation}
	or,
	\begin{equation}\label{eq:reladd1}
		\chi(\varphi_{\bf e}({\bf s}))=\frac{1}{2}\begin{pmatrix}
			1 \\ {\bf v_e} \oplus {\bf v_s}
		\end{pmatrix} \in \mathcal S(\spinf),
	\end{equation}
	and so
	\begin{equation}\label{eq:reladd2}
		\chi(\psi_{\bf e}({\bf s}))=e_0(1+{\bf v}_{\bf e}\cdot {\bf v}_{\bf s}) \, \frac{1}{2}\begin{pmatrix}1 \\{\bf v}_{\bf e}\oplus{\bf v}_{\bf s}\end{pmatrix} \in \tilde{\mathcal S}(\spinf),
	\end{equation}
	where the relativistic sum ${\bf v}_{\bf e}\oplus {\bf v}_{\bf s}$ is defined as follows: if $\|{\bf v_e}\|<1$, then
	\begin{equation}\label{eq:reladd}
		{\bf v_e}\oplus{\bf v_s}:={1\over 1+{\bf v_e}\cdot{\bf v_s}}\left\{{\bf v_e}+{1\over \gamma_{\bf v_e}}{\bf v_s}+{\gamma_{\bf v_e}\over 1+\gamma_{\bf v_e}}({\bf v_e}\cdot{\bf v_s}){\bf v_e}\right\},  
	\end{equation}
	$\gamma_{\bf v_e}$ is the Lorentz factor defined by
	\begin{equation}
		\gamma_{\bf v_e}:={1\over \sqrt{1-\Vert {\bf v_e}\Vert^2}},
	\end{equation}
	and, if $\|{\bf v_e}\|=1$,
	\beq\label{eq:reladdunitvector}
	{\bf v_e}\oplus{\bf v_s}:={\bf v_e}.
	\eeq
	Apart from the case of collinear vectors, the composition of Lüders operations is neither associative nor commutative due to the action of the so-called \textit{Thomas gyration operator}, see \cite{Ungar:08} for more details. This is particularly important to keep in mind when we write the expression of a post-measurement generalized state issued by a sequential Lüders operation as the following:
	\beq\label{eq:compeffects}
	\chi(\psi_{{\bf e}^2}(\psi_{{\bf e}^1}(s_0{\bf s})))=e^1_0e_0^2s_0(1+{\bf v}_{{\bf e}^1}\cdot{\bf v}_{\bf s})(1+{\bf v}_{{\bf e}^2}\cdot ({\bf v}_{{\bf e}^1} \oplus {\bf v}_{\bf s})) \frac{1}{2}\begin{pmatrix}
		1 \\ {\bf v}_{{\bf e}^2}\oplus ({\bf v}_{{\bf e}^1}\oplus {\bf v}_{\bf s})\end{pmatrix}.
	\eeq

	We end by recalling the fundamental chromatic matching equation, that will be applied in section \ref{sec:lcderiv}  to obtain the characterization of lightness constancy. From \cite{BerthierProvenzi:2022PRS} we have the following result: given two couples of chromatic states-effects $({\bf s}^1,{\bf e}^1)$ and $({\bf s}^2,{\bf e}^2)$, the equation
	\begin{equation}\label{eq:chromatch1}
		\varphi_{{\bf e}^1}({\bf s}^1)=\varphi_{{\bf e}^2}({\bf s}^2),
	\end{equation}
	or, equivalently, 
	\begin{equation}\label{eq:chromatch2}
		{\bf v}_{{\bf e}^1}\oplus{\bf v}_{{\bf s}^1}={\bf v}_{{\bf e}^2}\oplus{\bf v}_{{\bf s}^2},
	\end{equation}
	represents the \textit{chromatic matching equation} between $({\bf s}^1,{\bf e}^1)$ and $({\bf s}^2,{\bf e}^2)$ that establishes the perception of the same  chromatic information.

	\subsubsection{A practical application: white balance of color images}
	In this subsection we are going to mention a first application of the concepts introduced above to color image processing. In particular, we are going to explain how a L\"uders operation can be used as a chromatic adaptation transform for white balance. We underline that our aim here is only to provide a concrete example of the practical usefulness of the theory recalled in this section, a thorough comparison with other chromatic adaptation transforms is out of the scope of our paper and it will be coherently treated in a future work.
	
	White balance algorithms are meant to emulate the capability of the human visual system to adapt to non-neutral illumination conditions. More precisely they consist of two steps:
	\begin{itemize}
		\item an illuminant estimation algorithm, that identifies the illuminant(s) in the scene associating to it (each of them) a 3-dimensional vector ${\bf e}$;
		\item a Chromatic Adaptation Transform (CAT), parametrized by ${\bf e}$ that eliminates the presence of the illuminant returning an image as if the scene was lit by a neutral illuminant.
	\end{itemize}

	\noindent The first one to propose an association, although only heuristically justified, between chromatic adaptation and Lorentz boosts was H. Yilmaz in \cite{Yilmaz:62} and, to the best of our knowledge, the first Lorentz boost CAT was described in \cite{Guennec:21}.

	
	

	In \cite{BerthierProvenzi:2022PRS} the presence of Lorentz boosts in the quantum-like model was mathematically justified by proving that eq. \eqref{eq:reladd2}, i.e. the L\"uders operation relative to an effect ${\bf e}$, can be re-written in the following way:
	\begin{equation}\label{eq:LudBst}
		\chi(\psi_{\bf e}({\bf s}))=\frac{e_0}{\gamma_{{\bf v}_{\bf e}}}B({\bf v}_{\bf e}) \frac{1}{2}\begin{pmatrix}1 \\{\bf v}_{\bf s}\end{pmatrix} \equiv B_N({\bf e}) \frac{1}{2}\begin{pmatrix}1 \\{\bf v}_{\bf s}\end{pmatrix},
	\end{equation}
	where $B({\bf v}_{\bf e})$ is the Lorentz boost associated to the chromatic vector ${\bf v}_{\bf e}$, whose associated matrix is:
	\begin{equation}\label{eq:LudBstmatrix}
		[B({\bf v}_{\bf e})]= \begin{pmatrix}\gamma_{{\bf v}_{\bf e}} & \gamma_{{\bf v}_{\bf e}}{\bf v}_{\bf e}^t \\ \gamma_{{\bf v}_{\bf e}}{\bf v}_{\bf e} &\sigma_0 + \frac{\gamma_{{\bf v}_{\bf e}}^2}{1+\gamma_{{\bf v}_{\bf e}}}{\bf v}_{\bf e}{\bf v}_{\bf e}^t\end{pmatrix},
	\end{equation}
	and where $B_N({\bf e}):=\frac{e_0}{\gamma_{{\bf v}_{\bf e}}}B({\bf v}_{\bf e})$ is the normalized Lorentz boost associated to ${\bf e}$.
	
	The normalized Lorentz boost CAT has been implemented in a modified version of the classic HCV color space encoding Hering's opponent mechanism\footnote{In particular we modified just the H coordinate via simple interpolation techniques in order to recover the red-green opponency, which is absent in HCV.}. We represented both the image and the estimated illuminant ${\bf e}$ in this color domain and applied $B_N({\bf e})^{-1}$ to eliminate the presence of the color cast due to the illuminant. Clearly, since $B({\bf v}_{\bf e})^{-1}=B(-{\bf v}_{\bf e})$, we have that $B_N({\bf e})^{-1}= \frac{\gamma_{{\bf v}_{\bf e}}}{e_0}B(-{\bf v}_{\bf e})$.

	Figure \ref{fig:bureau} shows a stand-alone result of the version of the normalized Lorentz boost CAT just described, while Figure \ref{fig:panko} offers a visual comparison with the classical von Kries CAT.

	\begin{figure}[!ht]
		\minipage{0.49\textwidth}
		\includegraphics[width=\linewidth]{./images/input.jpg}
		\endminipage\hfill
		\minipage{0.49\textwidth}
		\includegraphics[width=\linewidth]{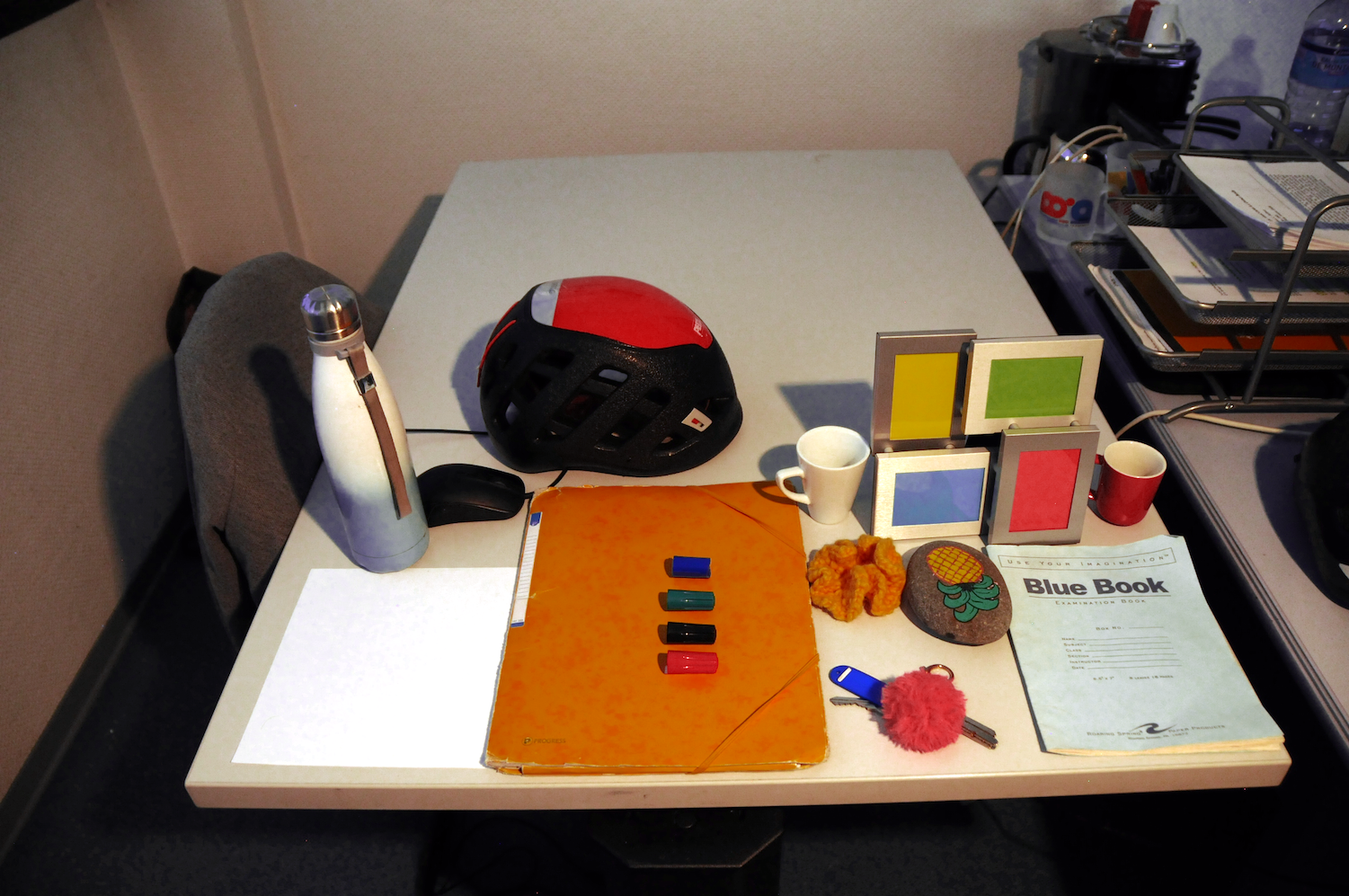}
		\endminipage\hfill
		\caption{\textit{Left}: input image. \textit{Right}: output image after white balance using the normalized Lorentz boost CAT. The illuminant has been estimated on the sheet of white paper placed on the left corner of the desk.}
		\label{fig:bureau}
	\end{figure}

	\begin{figure}[htbp]
		\minipage{0.32\textwidth}
		\includegraphics[width=\linewidth, scale = 0.5]{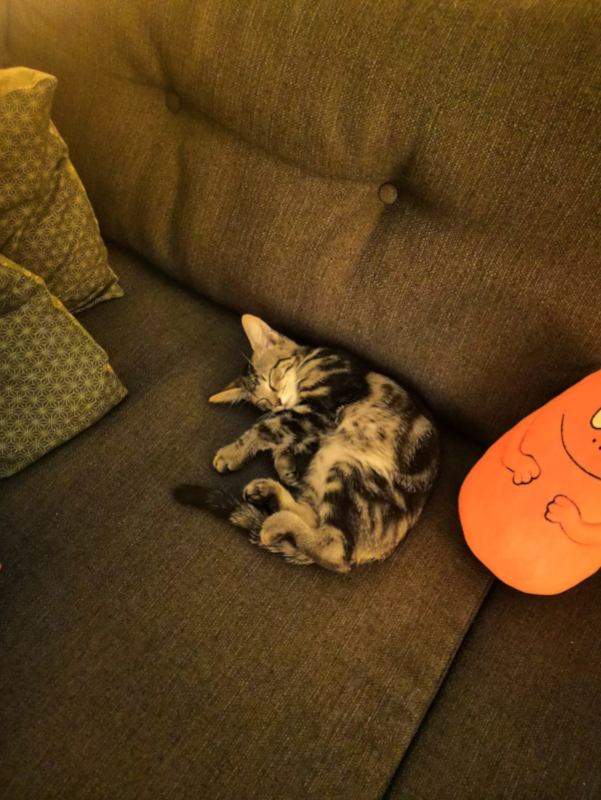}
		\endminipage\hfill
		\minipage{0.32\textwidth}
		\includegraphics[width=\linewidth]{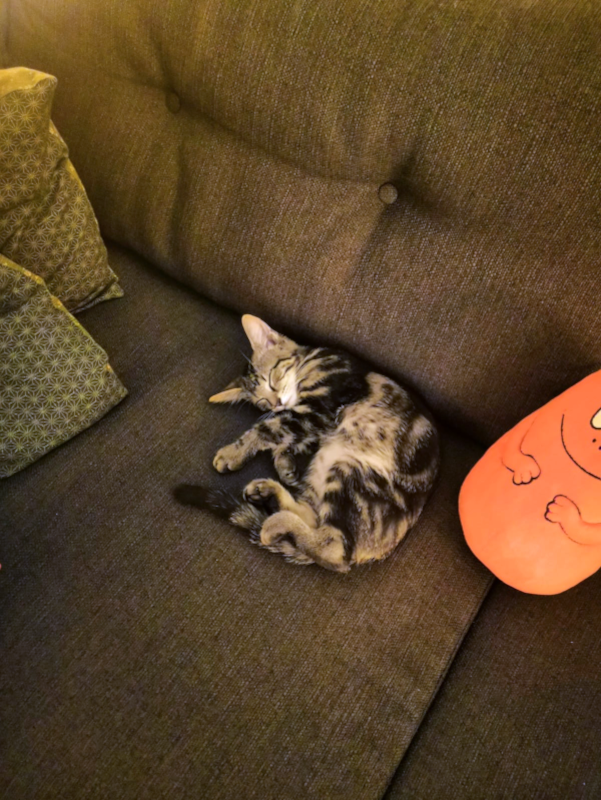}
		\endminipage\hfill
		\minipage{0.32\textwidth}%
		\includegraphics[width=\linewidth]{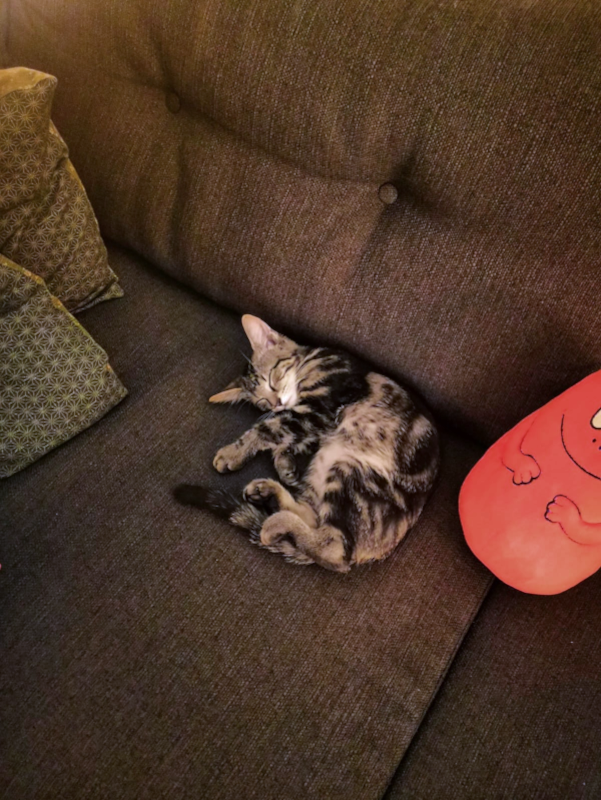}
		\endminipage
		\caption{\textit{Left}: input image. \textit{Center}: output image after white balance using the von Kries CAT. \textit{Right}: output image after white balance using the normalized Lorentz boost CAT. The white balanced images have been obtained using the same illuminant estimation.}
		\label{fig:panko}
	\end{figure}

	\bigskip

	\noindent Equipped with the knowledge about the quantum-like model of color perception that we have recalled in this section, we are ready to rigorously define the color perception attributes and to analyze the remarkable consequences of our definitions. For the sake of a clearer exposition, we shall divide this analysis into separate sections ordered following an increasing level of complexity.
	
	\section{The basic definitions: observer, illuminant, perceptual patch and perceived color from emitted and reflected light}\label{sec:obsillpercolor}
	
	In this section we provide the formalization of the most basic entities of our color perception theory. The modeling rules that we will follow are listed below: 
	\begin{itemize}
		\item any quantity whose chromatic features manifest themselves fused together with a normalized scalar factor will be described through a \textit{generalized state};
		\item any act of (physical or perceptual) color measurement and the (physical or perceptual) medium used to perform it will be associated to an \textit{effect};
		\item the measurement outcome will be identified with the \textit{post-measurement generalized state} induced by the action of the effect.
	\end{itemize}
	Our formalization starts with this very simple remark: \textit{a perceived color is the result of the measurement of a physical color stimulus performed by the visual system of a human observer}. 
	
	This means that a human observer is the medium through which a perceptual color measurement takes place, for this reason we model it as an effect.
	
	\bd[Observer]
	\em An observer $o$ measuring a color stimulus is identified with an effect ${\bf o}=(o_0,{\bf v_o}) \in \mathcal{E}$, $o_0\in [0,1]$ and ${\bf v_o} \in \cal D$. 
	\ed 
	
	The color stimulus hitting the eyes of $o$ can be either a light \textit{emitted} by a source of radiation or a light \textit{reflected} from the patch of a surface lit by an illuminant. Let us first formalize the former situation.
	
	\bd[Emitted light stimulus] 
	\em An emitted light stimulus $\ell$ is identified with the generalized state $\ell_0 \bell$, $\ell_0\in [0,1]$ and $\bell \in \mathcal S(\hr)$. The real quantity $\ell_0$ is the normalized light intensity and $\bell$ carries the intrinsic chromatic features.
	\ed 
	
	\bd[Achromatic and white light]
	An achromatic light is an emitted light stimulus with $\ell_0\in [0,1]$ and $\bell = \bf s_a$. If, in particular, $\ell_0=1$, then we call it a white light and we write $\bell^W=\bf s_a$.
	\ed 
	
	The act of measuring an emitted light stimulus $\ell$ by an observer $o$ produces a perceived color through the Lüders operation associated to the effect $\bf o$.
	
	\bd[Perceived color from a light stimulus]\label{def:colorfromlight}
	\em Given an observer $o$ and an emitted light stimulus $\ell$, i.e. the couple $(\bf{o},\ell_0 \bell)$, the color perceived by $o$ from $\ell$ is the post-measurement generalized state $\psi_{\bf o}(\ell_0 \bell)\in \tilde{\mathcal S}(\hr)$.
	\ed 
	Notice that this definition is coherent with the three-dimensional nature of perceived colors, in fact eq. \eqref{eq:varphi} implies:
	\beq\label{eq:oexplicit} 
	\psi_{\bf o}(\ell_0 \bell) = o_0 \ell_0(1+{\bf v_o} \cdot {\bf v_{\bell}}) \, \varphi_{\bf o}(\bell) = \langle {\bf o} \rangle_{\ell_0 \bell} \  \varphi_{\bf o}(\bell), 
	\eeq 
	with $\langle {\bf o} \rangle_{\ell_0 \bell}\in  [0,1]$ and $\varphi_{\bf o}(\bell) \in \mathcal S(\hr)$.
	
	Thanks to eq. \eqref{eq:achreffectstate}, we know that if an observer $o_{\text a}$ is associated to an achromatic effect $\bf o_a$, then
	\beq
	\varphi_{\bf o_a}({\bell}) = \bf \bell,
	\eeq 
	which means that the chromatic state of the color perceived by $o_{\text a}$ from the light source $\ell =\ell_0 \bell$ is exactly its intrinsic chromatic state $\bell$. 
	
	
	Formula \eqref{eq:oexplicit} shows explicitly the role played by the effect magnitude $o_0$ and by the effect vector $\bf v_o$: $o_0$ describes \textit{how the observer perceives the intensity} of the color stimulus, while $\bf v_o$ describes \textit{the adaptation state of the observer}.
	
	\smallskip 
	
	\noindent Let us now turn our attention to color stimuli from non-emitting surfaces. While the perceptual measurement of an emitted light stimulus consists simply in the act of observing it, a non-emitting surface needs an additional step: before being observed, it must be illuminated. For this reason, the formalization of the concept of perceived color from a reflected light requires the preliminary definition of illuminant. Being the medium that permits to perform a measurement process, an illuminant is identified with an effect\footnote{An observer $o$ can be thought as a \textit{perceptual effect}, an illuminant $o$ can be interpreted as a \textit{physical effect}.}.
	
	\bd[Illuminant and achromatic illuminant]
	\em An illuminant $\iota$ needed to light up a non-emitting surface in order to measure its color is identified with an effect $\biota = (\iota_0, {\bf v}_{\biota})$, $\iota_0\in [0,1]$, $\bf v_{\biota}\in \cal D$. The real quantity $\iota_0$ represents the illuminant intensity, while ${\bf v}_{\biota}$ carries the chromatic features. If ${\bf v}_{\biota}=\bf 0$, $\iota$ is called an achromatic illuminant. 
	\ed 
	Now let us pass to the definition of patch (or \textit{area}) of a non-emitting surface. Without being illuminated, a surface patch is characterized only by its intrinsic properties that establish how much light the surface reflects and how it interacts with the different spectral components of the incoming radiation. These features are fused together, which motivates the next definition.
	
	\bd[Patch]
	\em The patch $p$ of a non-emitting surface is identified with a generalized state $p_0{\bf p}$, $p_0\in [0,1]$ and ${\bf p}\in \mathcal S(\hr)$. The real quantity $p_0$ represents the overall proportion of the illuminant intensity that $p$ is able to reflect and $\bf p$ carries the intrinsic chromatic properties of $p$.
	\ed
	
	\begin{definition}[Achromatic and white patch]
		A patch $p=p_0 \bf p$ with ${\bf p}=\bf s_a$ is called achromatic. In particular, if $p_0=1$, then we call it white patch and we write ${\bf p}^W=\bf s_a$.
	\end{definition}
	
	When a patch is lit by an illuminant $\iota$ it can be observed, becoming a perceptual patch, as defined below. 
	
	\bd[Perceptual patch]
	\em A perceptual patch $r$ is a post-measurement generalized state\footnote{The letter $r$ reminds the fact that the generalized state $r_0\bf r$ is issued by the light reflected by $p$.}  $r_0\bf r$, $r_0\in [0,1]$, ${\bf r}\in \mathcal{S}(\hr)$, given by a physical patch $p$ lit by an illuminant $\iota$, i.e. $r_0{\bf r}=\psi_{\biota}(p_0{\bf p})$.
	\ed
	This definition is the \textit{perceptual} counterpart of the well-known \textit{physical} formula 
	\beq\label{eq:imageformation}
	I_p(\lambda,x)=L(\lambda)R_p(\lambda,x), 
	\eeq 
	typically used in image formation models, see e.g. \cite{Gevers:12,Provenzi:2017}. $I_p(\lambda,x)$ is the image information about the physical patch $p$ that has been acquired by a spectrophotometer at the wavelength $\lambda$ and at the spatial position $x$, $L(\lambda)$ is the luminance of the radiation used to light up the material (supposed to be spatially uniform, which explains the absence of the variable $x$) and $R_p(\lambda,x)$ is the patch reflectance at the wavelength $\lambda$ and at the point $x$. When $I_p(\lambda,x)$ is acquired, the data about $L$ and  $R$ are fused together.
	
	We are now ready to give the definition of perceived color of a patch.
	
	\bd[Perceived color from an illuminated patch]
	\em Given an observer $o$, a surface patch $p$ and an illuminant $\iota$, i.e. the triple $({\bf o},{\biota},p_0{\bf p})$, the color perceived by $o$ from the perceptual patch $r=\psi_{\biota}(p_0{\bf p})$ is the post-measurement generalized state $\psi_{\bf o}(r)=\psi_{\bf o}(\psi_{\biota}(p_0{\bf p}))\in \tilde{\mathcal S}(\hr)$.
	\ed 
	We can interpret the sequential operation $\psi_{\bf o}\circ \psi_{\biota}$ obtained via the combined action of the (physical) effect $\biota$ and the (perceptual) effect $\bf o$ as a Lüders operation associated to a single (perceptual) effect $\tilde{\bf{o}}$ defined either by the equation
	\beq\label{eq:combinedeffect}
	\psi_{\tilde{\bf{o}}}(p_0{\bf p}):=(\psi_{\bf o}\circ \psi_{\biota})(p_0{\bf p})=\psi_{\bf o}(r),
	\eeq 
	or, thanks to eq. \eqref{eq:compeffects}, by the more explicit formula
	\beq\label{eq:otilde}
	\chi(\psi_{\tilde{\bf{o}}}(p_0{\bf p}))=o_0\iota_0p_0(1+{\bf v}_{{\biota}}\cdot{\bf v}_{\bf p})(1+{\bf v}_{{\bf o}}\cdot ({\bf v}_{{\biota}} \oplus {\bf v}_{\bf p})) \frac{1}{2}\begin{pmatrix}
		1 \\ {\bf v}_{{\bf o}}\oplus ({\bf v}_{{\biota}}\oplus {\bf v}_{\bf p})\end{pmatrix}.
	\eeq 
	Thanks to eqs. \eqref{eq:psiphiachr} and \eqref{eq:achreffectstate}, if $\iota$ is an achromatic illuminant $\biota_{\bf a}=(\iota_0,{\bf 0})$ we have
	\beq
	r_{\text a} := \psi_{\biota_{\bf a}}(p_0{\bf p})=\iota_0 p_0 \varphi_{\biota_{\bf a}}({\bf p})= \iota_0 p_0 \bf p,
	\eeq 
	If, moreover, the observer $o$ is represented by an achromatic effect $\bf o_a$, then 
	\beq\label{eq:odapted}
	\varphi_{\bf o_a}(r_{\text{a}})=\varphi_{\bf o_a}(\psi_{\biota_{\bf a}}(p_0{\bf p}))=\varphi_{\bf o_a}(\iota_0 p_0{\bf p})=\bf p,
	\eeq 
	thus, \textit{such an observer perceives the chromatic state of a physical patch lit by an achromatic illuminant as it is}.
	
	Let us conclude this section with two remarks. The first is that the concepts of color perceived by an observer from an emitted light stimulus, see def. \ref{def:colorfromlight}, and from an illuminated surface patch, see eq.  \eqref{eq:combinedeffect}, in spite of having different interpretations, can be  characterized by \textit{the same mathematical object}: a post-measurement generalized state. For this reason, hereinafter, when it is not meaningful to distinguish between the two cases, we will deal with a perceived color by using the abstract and unifying notation represented by $\psi_{\bf e}(s_0{\bf s})$.
	
	
	The second remark refers to the link between two apparently different definitions of perceived colors that we have done: in section \ref{subsubsec:colorsolid} a perceived color has been defined as a an effect, while in the present section we have identified it with a post-measurement generalized state induced by an effect. In fact, if ${\bf e}:=(e_0,{\bf v_e})$ is an effect, then one can associate to ${\bf e}$ the perceived color $\psi_{\bf e}({\bf s_a})=e_0\varphi_{\bf e}({\bf s_a})$, this correspondence being clearly one-to-one and onto.

	\section{Definition of the achromatic attributes: brightness and lightness}\label{sec:lightBright}
	Defining a meaningful terminology to describe the achromatic component of a perceived color is a delicate issue. The title of \cite{Kingdom:2011} emblematically refers to it as an \textit{unrelenting controversy}. This confusion is particularly evident when one reads names as lightness, brightness, luminance, luma, value or intensity used as synonyms to describe the achromatic attribute in image processing.
	
	In this section we will provide a mathematically rigorous proposal for the definitions of brightness and lightness. To motivate our proposals, we start by reporting the following two descriptions that refer to the case of light reflected by a physical patch lit by an illuminant. 
	
	Quoting \cite{Gilchrist:07}: `\textit{the physical counterpart of lightness is the permanent property of a surface that determines what percentage of light the surface reflects. Surfaces that appear white reflect about $90\%$ of the light striking them. Black surfaces reflect about $3\%$. In short, lightness is perceived reflectance}'.
	
	Quoting \cite{Kingdom:2011} : `\textit{the physical counterpart of brightness is called luminance, that is, the absolute intensity of light reflected in the direction of the observer’s eye by a surface (or at least coming from a certain part of the visual field). In short, if lightness is perceived reflectance, brightness is perceived luminance. The reflectance of an object is a relatively permanent property, whereas its luminance is transient}'.
	
	The basic  information brought by the references quoted above is that, in order to extract brightness and lightness from the perceived color $\psi_{\bf e}(s_0{\bf s})$, we must be able to meaningfully extract a percentage out of it which has to verify suitable perceptual robustness properties.
	
	By the fact that $s_0\in [0,1]$ and thanks to eqs. \eqref{eq:constraint01} and  \eqref{eq:psigen}, the expected value of an effect $\bf e$ on the generalized state $s_0{\bf s}$, i.e. the trace of $\psi_{\bf e}(s_0{\bf s})$, belongs to the interval $ [0,1]$. This is the most natural way to associate a percentage to $\psi_{\bf e}(s_0{\bf s})$ and it leads to our proposal for the definition of brightness. 
	
	\begin{definition}[Brightness of a perceived color from an emitted light]\label{def:brightlight}
		\em Given an observer $o$, ${\bf o}=(o_0,{\bf v_o})$, the brightness of the color $\psi_{\bf o}(\ell_0{\bell})$ perceived by $o$ from an emitted light stimulus $\ell_0 \bell$ is given by
		\begin{equation}\label{eq:brbr}
			{\mathcal B}(\psi_{\bf o}(\ell_0{\bell})):={\Tr}(\psi_{\bf o}(\ell_0{\bell}))=o_0\ell_0(1+{\bf v}_{\bf o}\cdot {\bf v}_{\bell}).
		\end{equation}
	\end{definition}
	The following result is immediate, we state it for white light because we need it to define lightness, but it can be extended to an arbitrary achromatic emitted light by replacing $o_0$ with $\ell_0o_0$, $\ell_0\in [0,1]$.
	\begin{proposition}[Robustness of the white light brightness]\label{prop:roblightbright}
		Given any observer ${\bf o}=(o_0,{\bf v_o})$, the brightness perceived by $o$ from the white light $\bell^W$ is:
		\begin{equation}\label{eq:brightwlight}
			{\mathcal B}(\psi_{\bf o}(\bell^W))=o_0,
		\end{equation}
		so the brightness of the white light does not depend on the effect vector of $o$.
	\end{proposition}
	Now we treat the case of reflected light.
	
	\begin{definition}[Brightness of a perceived color from a reflected light]\label{def:brightrefl}
		\em Given a couple observer-illuminant $(o,\iota)$, ${\bf o}=(o_0,{\bf v_o})$, ${\biota}=(\iota_0,{\bf v_{\biota}})$, the brightness of the color $\psi_{\bf o} (\psi_{\biota}(p_0{\bf p}))$ perceived by $o$ from a patch $p_o{\bf p}$ lit by $\iota$ is:
		\begin{equation}\label{eq:brightrefl}
			{\mathcal B}(\psi_{\bf o} (\psi_{\biota}(p_0{\bf p})) = o_0\iota_0p_0(1+{\bf v}_{{\biota}}\cdot{\bf v}_{\bf p})(1+{\bf v}_{{\bf o}}\cdot ({\bf v}_{{\biota}} \oplus {\bf v}_{\bf p})).
		\end{equation}
	\end{definition}
	The equivalent of proposition \ref{prop:roblightbright} in the case of reflected light is the following result, which can be extended to achromatic patches by replacing $o_0\iota_0$ with $o_0\iota_0p_0$.
	
	\begin{proposition}[Robustness of white patch brightness under achromatic illuminant]
		Given a couple observer-illuminant $(o,\iota)$, ${\bf o}=(o_0,{\bf v_o})$, ${\biota}=(\iota_0,{\bf v_{\biota}})$, the brightness perceived by $o$ from the white patch ${\bf p}^W$ lit by $\iota$ is:
		\begin{equation}\label{eq:brightwpatch}
			{\mathcal B}(\psi_{\bf o} (\psi_{\biota}({\bf p}^W)) = o_0\iota_0(1+{\bf v_o} \cdot {\bf v_{\biota}}),
		\end{equation}
		hence, the brightness of the white patch does not depend on the effect vector of $o$ if and only if $\iota$ is an achromatic illuminant $\iota_a$, in which case we have:
		\begin{equation}
			{\mathcal B}(\psi_{\bf o} (\psi_{\biota_{\bf p}}({\bf p}^W)) = o_0\iota_0.
		\end{equation}
	\end{proposition}
	Let us now pass to the definition of lightness. The following reasoning will give a more substantiated basis to the intuitive eq. \eqref{eq:light} proposed in \cite{Fairchild:13}.
	
	An observer cannot distinguish an isolated chromatic patch lit by an achromatic illuminant from an achromatic one lit by a chromatic illuminant. The physical counterpart of this statement is the impossibility of recovering the reflectance $R_p(\lambda,x)$ from the sole knowledge of $I_p(\lambda,x)$ in  formula \eqref{eq:imageformation}: it is clear that, without any further hypothesis on $R_p(\lambda,x)$, or on the luminance $L(\lambda)$, this problem is ill-posed.
	
	This is the reason why several hypotheses, e.g. white patch, gray world, gray edge and so on, have been formulated in order to solve this problem, see e.g. \cite{Gevers:12,Provenzi:2017} for an overview. Among them, the only hypothesis that can be meaningfully applied to unrelated colors is the white patch (because unrelated colors, by definition, do not have a surround), i.e. the physical assumption that there exists a patch $p^W$, among those observed under the same illuminant, that has perfect reflectance, i.e. such that $R_{p^W}(\lambda,x)\equiv 1$.
	
	If this hypothesis is satisfied, then formula \eqref{eq:imageformation} gives $I_{p^W}(\lambda,x)=L(\lambda)$, i.e. the image information acquired from the white patch $p^W$ agrees with the luminance of the illuminant, hence we can retrieve the reflectance of each patch $p$ from the image information $I_{p}(\lambda,x)$ simply dividing it by $I_{p^W}(\lambda,x)$, i.e.
	\beq
	R_p(\lambda,x)=\frac{I_{p}(\lambda,x)}{I_{p^W}(\lambda,x)}.
	\eeq 
	As before, we distinguish our definition of lightness for emitted and reflected light, starting by the former case.
	
	\begin{definition}[Lightness of a perceived color from an emitted light]\label{def:lightemitted}
		\em Given an observer $o$, ${\bf o}=(o_0,{\bf v_o})$, the lightness of the color $\psi_{\bf o}(\ell_0{\bell})$ perceived by $o$ from an emitted light stimulus $\ell_0 \bell$ is given by the ratio between its brightness, eq. \eqref{eq:brbr}, and the brightness of the white light, eq. \eqref{eq:brightwlight}, i.e.
		\begin{equation}
			{\mathcal L}(\psi_{\bf o}(\ell_0{\bell})):=\frac{{\mathcal B}(\psi_{\bf o}(\ell_0{\bell}))}{{\mathcal B}(\psi_{\bf o}({\bell}^W))}=\ell_0(1+{\bf v}_{\bf o}\cdot {\bf v}_{\bell}).
		\end{equation}
		
	\end{definition}
	The lightness perceived from an achromatic emitted light coincides with its normalized intensity $\ell_0$ independently of the observer:
	\beq
	{\mathcal L}(\psi_{\bf o} (\ell_0{\bf s_a}))=\ell_0, \quad \forall {\bf o}.
	\eeq 
	In particular, the lightness of the white light is normalized to 1. 
	
	When ${\bf v}_{\bf o}=\bf 0$, the lightness of any color perceived from an emitted light coincides with the light intensity independently of the chromatic state of the emitted light:
	\beq
	{\mathcal L}(\psi_{\bf o_a} (\ell_0\bell))=\ell_0, \quad \forall \bell.
	\eeq 
	
	\begin{definition}[Lightness of a perceived color from a reflected light]\label{def:lightreflected}
		\em Given a couple observer-illuminant $(o,\iota)$, ${\bf o}=(o_0,{\bf v_o})$, ${\biota}=(\iota_0,{\bf v_{\biota}})$, the lightness of the color $\psi_{\bf o} (\psi_{\biota}(p_0{\bf p}))$ perceived by $o$ from the patch $p_0\bf p$ lit by $\iota$ is given by the ratio between its brightness, eq. \eqref{eq:brightrefl}, and the brightness of the white patch lit by the same illuminant $\iota$, eq.  \eqref{eq:brightwpatch}, i.e. 
		\begin{equation}
			{\mathcal L}(\psi_{\bf o} (\psi_{\biota}(p_0{\bf p})):=\frac{{\mathcal B}(\psi_{\bf o} (\psi_{\biota}(p_0{\bf p}))}{{\mathcal B}(\psi_{\bf o} (\psi_{\biota}({\bf p}^W))}=p_0\frac{(1+{\bf v}_{{\biota}}\cdot{\bf v}_{\bf p})(1+{\bf v}_{{\bf o}}\cdot ({\bf v}_{{\biota}} \oplus {\bf v}_{\bf p}))}{1+{\bf v}_{\bf o}\cdot{\bf v}_{\biota}}.
		\end{equation}
	\end{definition}
	Notice that the lightness of an achromatic patch coincides with $p_0$, the overall percentage of illuminant intensity that the patch is able to reflect, regardless of the chromaticity of the illuminant $\iota$ and the effect vector of the observer $o$: 
	\beq
	{\mathcal L}(\psi_{\bf o} (\psi_{\biota}(p_0{\bf s_a}))=p_0, \quad \forall ({\bf o},\biota).
	\eeq 
	In particular, the lightness of the white patch is normalized to 1. 
	
	Differently from the case of emitted light, the lightness of a surface color perceived by an observer with ${\bf v}_{\bf o}=\bf 0$ is not simply $p_0$ but
	\beq
	{\mathcal L}(\psi_{\bf o_a} (\psi_{\biota}(p_0{\bf p}))=p_0(1+{\bf v}_{{\biota}}\cdot{\bf v}_{\bf p}),
	\eeq 
	this quantity reduces to $p_0$ when $\iota$ is an achromatic illuminant:
	\beq
	{\mathcal L}(\psi_{\bf o_a} (\psi_{\biota_{\bf p}}(p_0{\bf p}))=p_0.
	\eeq 
	As a final remark, we notice that the fact that brightness and lightness differ by the multiplicative constant represented by the brightness of the perceived white area implies that a logarithmic variation in brightness is equal to a logarithmic variation in lightness, as in the Weber-Fechner's law, see e.g. \cite{Goldstein:13}. This might justify the choice of $ds/s$, which is invariant under scalar multiplication of $s$, as metric for the achromatic component of a color.

	\section{Definition of perceptual chromatic attributes: colorfulness, saturation, chroma and hue}\label{sec:chromattributes}
	In this section we discuss two possible ways of defining the chromatic attributes of colorfulness, chroma and saturation. In the first subsection, we use chromatic opponency to characterize them, while in the second subsection we employ the concept of relative quantum entropy, that will also be used in a third subsection to define the concept of hue. 
	
	\subsection{Euclidean definition of colorfulness, saturation and chroma of a perceived color via chromatic opponency}\label{subsec:chromopponency}
	
	The expectation values of the real Pauli matrices $\sigma_1,\sigma_2$ on a chromatic state $\bf s$ provide the degrees of opponency that characterize the chromatic perception of the perceived color $\psi_{\bf e}(s_0\bf s)$. 
	
	Our aim here is to define the chromatic attributes of \textit{colorfulness}, \textit{chroma} and \textit{saturation} using only the information about chromatic opponency. 
	
	Moreover, we want to translate into rigorous equations the intuitive formulae \eqref{eq:Saturation} and \eqref{eq:Chroma}, that we recall here:
	\beq
	\text{Saturation}=\frac{\text{Colorfulness}}{\text{Brightness}} \quad \text{and} \quad \text{Chroma}=\frac{\text{Colorfulness}}{\text{Brightness(White)}}.
	\eeq 
	Since we have already defined the concepts of brightness, what remains to be defined is just the colorfulness. In fact, given the perceived color $\psi_{\bf e}(s_0{\bf s})$, if we know how to define its colorfulness $\text{Col}(\psi_{\bf e}(s_0{\bf s}))$, then its saturation `Sat' and chroma `Chr' are, respectively,
	\begin{equation}\label{eq:linearitychromasaturation}
		{\rm Sat(\psi_{\bf e}(s_0{\bf s}))}=\frac{{\rm Col}({\psi_{\bf e}(s_0{\bf s})})}{{\mathcal B}({\psi_{\bf e}(s_0{\bf s})})} \quad \text{and} \quad {\rm Chr(\psi_{\bf e}(s_0{\bf s}))}=\frac{{\rm Col}({\psi_{\bf e}(s_0{\bf s})})}{{\mathcal B}(\psi_{\bf e}({\bf s_a}))}.
	\end{equation}
	Alternatively, if we knew how to define saturation, we could define colorfulness by inverting the linear relationships in eq. \eqref{eq:linearitychromasaturation}, i.e.
	\beq\label{eq:linlin01} 
	{\rm Col}({\psi_{\bf e}(s_0{\bf s})}) = \beta \, {\rm Sat(\psi_{\bf e}(s_0{\bf s}))}\quad \text{and} \quad  {\rm Chr}({\psi_{\bf e}(s_0{\bf s})}) = \lambda \, {\rm Sat(\psi_{\bf e}(s_0{\bf s}))},
	\eeq 
	where 
	\beq\label{eq:linlin02}
	\begin{cases}
		\beta:={\mathcal B}({\psi_{\bf e}(s_0{\bf s})})\\
		\lambda:={\mathcal L}({\psi_{\bf e}(s_0{\bf s})})={\mathcal B}({\psi_{\bf e}(s_0{\bf s})})/{{\mathcal B}(\psi_{\bf e}({\bf s_a}))}
	\end{cases}.
	\eeq
	We will exploit this remark in the next subsection.
	
	We start with a preliminary definition.
	
	\begin{definition}[$i$-th degrees of opponency of a perceived color]
		Let $\psi_{\bf e}(s_0{\bf s})$ be a perceived color. Then, for $i=1,2$, its:
		\begin{itemize}
			\item {\em $i$-th degree of colorfulness opponency} is 
			\begin{equation}
				{\rm Col}_i(\psi_{\bf e}(s_0{\bf s})):=\langle\sigma_i\rangle_{\psi_{\bf e}(s_0{\bf s})} \ ;
			\end{equation}
			\item {\em $i$-th degree of saturation opponency} is 
			\begin{equation}
				{\rm Sat}_i(\psi_{\bf e}(s_0{\bf s})):={\langle\sigma_i\rangle_{\psi_{\bf e}(s_0{\bf s})}\over {\rm Tr}(\psi_{\bf e}(s_0{\bf s}))} \ ;
			\end{equation}
			\item {\em $i$-th degree of chroma opponency} is 
			\begin{equation}
				{\rm Chr}_i(\psi_{\bf e}(s_0{\bf s})):={\langle\sigma_i\rangle_{\psi_{\bf e}(s_0{\bf s})}\over {\rm Tr}(\psi_{\bf e}({\bf s_a}))}.
			\end{equation}
		\end{itemize}
	\end{definition}
	
	Now we have to face the problem of suitably combine the $i$-th degree of opposition of these chromatic attributes in order to obtain a positive real number that defines the attribute itself. If we had to follow the Euclidean choice of classical colorimetry we would give the following definitions.
	
	\begin{definition}[Euclidean definitions of colorfulness, saturation and chroma of a perceived color]
		Given the perceived color $\psi_{\bf e}(s_0{\bf s})$, its:
		\begin{itemize}
			\item colorfulness is 
			\begin{equation}
				{\rm Col}(\psi_{\bf e}(s_0{\bf s}))=\sqrt{\left[{\rm Col}_1(\psi_{\bf e}(s_0{\bf s}))\right]^2+\left[{\rm Col}_2(\psi_{\bf e}(s_0{\bf s}))\right]^2} \ ;
			\end{equation}
			\item saturation is 
			\begin{equation}
				{\rm Sat}(\psi_{\bf e}(s_0{\bf s}))=\sqrt{\left[{\rm Sat}_1(\psi_{\bf e}(s_0{\bf s}))\right]^2+\left[{\rm Sat}_2(\psi_{\bf e}(s_0{\bf s}))\right]^2} \ ;
			\end{equation}
			\item chroma is 
			\begin{equation}
				{\rm Chr}(\psi_{\bf e}(s_0{\bf s}))=\sqrt{\left[{\rm Chr}_1(\psi_{\bf e}(s_0{\bf s}))\right]^2+\left[{\rm Chr}_2(\psi_{\bf e}(s_0{\bf s}))\right]^2}.
			\end{equation}
		\end{itemize}
	\end{definition}
	With such definitions the linear relations of eq. \eqref{eq:linearitychromasaturation} are satisfied. 
	
	We now pass to the discussion of a second possible way to define chromatic attributes that we deem more coherent with the quantum-like theory of color perception that lies at the basis of our work.
	
	\subsection{Definition of colorfulness, saturation and chroma of a perceived color via relative quantum entropy}\label{subsec:chromrelentropy}
	
	Here we propose an alternative description of the perceptual chromatic attributes based on the notion of relative (quantum) entropy, for more information about this concept we refer to \cite{Cortese:02,Auletta:09} and also to \cite{Ohya:04} or \cite{Audenaert:11} for the proofs of its properties that we shall quote here.
	
	Given two states $\bf s$ and $\bf t$ represented by the density matrices $\rho_{{\bf s}}$ and $\rho_{{\bf t}}$, the relative entropy between them is defined as the following real value:
	\begin{equation}\label{eq:relent}
		R(\rho_{{\bf s}}||\rho_{{\bf t}}):={\rm Tr}\left[ \rho_{{\bf s}}\log_2\rho_{{\bf s}}-\rho_{{\bf s}}\log_2\rho_{{\bf t}}\right].
	\end{equation}
	Actually, the so-called \textit{Klein inequality}, establishes a sort of  `definite positivity' for $R$ in the following sense: $R(\rho_{{\bf s}}||\rho_{{\bf t}})\ge 0$ for all $\rho_{{\bf s}}$ and $\rho_{{\bf t}}$ and $R(\rho_{{\bf s}}||\rho_{{\bf t}})=0$ if and only if $\rho_{{\bf s}}=\rho_{{\bf t}}$. 
	
	One of the most important reasons why we consider the relative entropy so inherently natural in our analysis of chromatic attributes in the quantum-like framework is that it can also be defined on generalized state density matrices. In fact, for all $\lambda>0$, $R$ satisfies the following sort of  `1-degree homogeneity':
	\begin{equation}\label{eq:lin}
		R(\lambda\rho_{{\bf s}}||\lambda\rho_{{\bf t}})=\lambda R(\rho_{{\bf s}}||\rho_{{\bf t}}).
	\end{equation}
	Unlike the von Neumann entropy, relative entropy `behaves well' with respect to scalar multiplication, it is thanks to this feature that we will be able to build a coherent system of linearly related definitions of saturation, chroma and colorfulness, which are the same quantity up to a scalar factor, as shown by the equations in formula \eqref{eq:linlin01}. 
	
	To obtain an explicit expression for $R$, let us consider two density matrices $\rho_{\bf s}$ and $\rho_{\bf t}$ with chromatic state vectors ${\bf v}_{\bf s}=(s_1,s_2)$ and ${\bf v}_{\bf t}=(t_1,t_2)$, respectively, i.e.
	\begin{equation}
		\rho_{\bf s}={1\over 2}\left(\begin{array}{cc}1+s_1 & s_2 \\ s_2 & 1-s_1 \end{array}\right), \quad \rho_{\bf t}={1\over 2}\left(\begin{array}{cc}1+t_1 & t_2 \\ t_2 & 1-t_1\end{array}\right).
	\end{equation}
	Let us also denote $r_{\bf s}:=\Vert{\bf v}_{\bf s}\Vert$, $r_{\bf t}:=\Vert{\bf v}_{\bf t}\Vert$, and $\cos \vartheta_{{\bf s},{\bf t}}:={\bf v}_{\bf s}\cdot{\bf v}_{\bf t}/r_{\bf s}r_{\bf t}$.  
	
	Technical computations lead to the following explicit expression:
	\begin{equation}\label{eq:Relentropy}
		\begin{split}
			R(\rho_{{\bf s}}||\rho_{{\bf t}}) & ={1\over 2}\log_2(1-r_{\bf s}^2)+{r_{\bf s}\over 2}\log_2\left(1+r_{\bf s}\over 1-r_{\bf s}\right)\\
			& -{1\over 2}\log_2(1-r_{\bf t}^2)-{r_{\bf s}\cos \vartheta_{{\bf s},{\bf t}}\over 2}\log_2\left(1+r_{\bf t}\over 1-r_{\bf t}\right).
		\end{split}
	\end{equation}
	As a particularly important case of eq. \eqref{eq:relent}, if ${\bf t}={\bf s_a}$, i.e. $\rho_{\bf t}=\rho_{\bf 0}$, the achromatic state, then $r_{\bf t}=0$ and:
	\begin{equation}
		R(\rho_{{\bf s}}||\rho_{\bf 0})={1\over 2}\log_2(1-r_{\bf s}^2)+{r_{\bf s}\over 2}\log_2\left(1+r_{\bf s}\over 1-r_{\bf s}\right)=\Sigma(r_{\bf s}),
	\end{equation}
	where, as in eq. \eqref{eq:satvonneumann} of section \ref{sec:recapquant}, $\Sigma(r_{\bf s})=1-S(\rho_{\bf s})$,  $S(\rho_{\bf s})$ being the von Neumann entropy of the state ${\bf s}$. 
	
	So, \textit{the relative entropy between $\rho_{{\bf s}}$ and the achromatic state agrees exactly with the definition of saturation proposed in} \cite{Berthier:2021JofImaging}, this is the strong argument that we have announced in section \ref{subsec:chromentropies} in favor of the interpretation of the von Neumann entropy, instead of the linear one, as a descriptor of chromatic purity. 
	
	Now, in order to define the saturation of the perceived color $\psi_{\bf e}(s_0{\bf s})$ we simply consider the chromatic state $\varphi_{\bf e}(s_0{\bf s})=\varphi_{\bf e}({\bf s})$ associated to it and we compute the relative entropy between its density matrix and $\rho_{\bf 0}$, as formalized in the following definition.
	
	\begin{definition}[Saturation of a perceived color]
		Given the perceived color $\psi_{\bf e}(s_0{\bf s})$, its saturation is
		\begin{equation}\label{eq:saturation}
			\begin{split}
				{\rm Sat}(\psi_{\bf e}(s_0{\bf s}))& =R(\rho_{\varphi_{\bf e}({\bf s})}||\rho_{\bf 0})=1-S(\rho_{\varphi_{\bf e}({\bf s})})= \Sigma (r_{\varphi_{\bf e}({\bf s})})\\
				& = {1\over 2}\log_2(1-r_{\varphi_{\bf e}({\bf s})}^2)+{r_{\varphi_{\bf e}({\bf s})}\over 2}\log_2\left(1+r_{\varphi_{\bf e}({\bf s})}\over 1-r_{\varphi_{\bf e}({\bf s})}\right),
			\end{split}
		\end{equation}
		with $r_{\varphi_{\bf e}({\bf s})}=\|{\bf v}_{\varphi_{\bf e}({\bf s})}\|$, where ${\bf v}_{\varphi_{\bf e}({\bf s})}=(\langle \sigma_1 \rangle_{\varphi_{\bf e}({\bf s})},\langle \sigma_2 \rangle_{\varphi_{\bf e}({\bf s})})$ is the chromatic state vector of $\varphi_{\bf e}({\bf s})$ which contains the intrinsic information about its degrees of chromatic opposition.
	\end{definition}
	Hence, ${\rm Sat}(\psi_{\bf e}(s_0{\bf s}))$ depends only on the effect $\bf e$ that permits the observation of the color and on its chromatic features, embedded in $r_{\varphi_{\bf e}({\bf s})}$, but \textit{not on} $s_0$.
	
	
	Instead, and crucially, if we define colorfulness and chroma following eqs. \eqref{eq:linlin01}, then the coefficient $s_0$ appears explicitly, as we can see next. 
	
	\begin{definition}[Colorfulness of a perceived color] 
		Given the perceived color $\psi_{\bf e}(s_0{\bf s})$ with brightness $\beta=\mathcal B(\psi_{\bf e}(s_0{\bf s}))=\Tr(\psi_{\bf e}(s_0{\bf s}))$, its \textit{colorfulness} is 
		\begin{equation}\label{eq:colorfulness}
			{\rm Col}(\psi_{\bf e}(s_0{\bf s}))=R(\beta \rho_{\varphi_{\bf e}(s_0{\bf s})}||\beta \rho_{{\bf s_a}})=\beta \, {\rm Sat}(\psi_{\bf e}(s_0{\bf s})).
		\end{equation}
	\end{definition}
	
	\begin{definition}[Chroma of a perceived color]
		Let $\psi_{\bf e}(s_0{\bf s})$ be a perceived color with lightness given by  $\lambda=\mathcal L(\psi_{\bf e}(s_0{\bf s}))=\Tr(\psi_{\bf e}(s_0{\bf s}))/\Tr(\psi_{\bf e}({\bf s_a}))$, then its \textit{chroma} is 
		\begin{equation}\label{eq:chroma}
			{\rm Chr}(\psi_{\bf e}(s_0{\bf s}))=R(\lambda  \rho_{\varphi_{\bf e}(s_0{\bf s})}||\lambda  \rho_{{\bf s_a}})=\lambda \, {\rm Sat}(\psi_{\bf e}(s_0{\bf s})).
		\end{equation}
	\end{definition}
	
	\subsection{Definition of hue of a perceived color via relative quantum entropy}\label{subsec:huerelentropy}
	As lightness and brightness, perceptual hue has a physical counterpart: the concept of {\em dominant wavelength} of a color stimulus. As presented in \cite{Wyszecky:82}, the dominant wavelength of a color stimulus is `{\em the wavelength of monochromatic stimulus that, when mixed with some specified achromatic stimulus, matches the given stimulus in color}'. In other words, the dominant wavelength characterizes any light mixture in terms of the monochromatic spectral light that elicits the same perception of hue. In the CIE chromaticity diagram, the dominant wavelength is the point of its border determined by the intersection with  the straight line that passes through the white point and the one associated to the given color. 
	
	In order to translate this concept within the quantum-like perceptual framework, motivated by the results of the previous subsection, we replace the concept of nearest Euclidean distance to the border of the CIE chromaticity diagram with that of minimal relative entropy between a given chromatic state and a pure state parameterized by a point of the border of $\mathcal D$.
	
	These considerations lead naturally to the following definition of hue.
	\begin{definition}[Hue of a perceived color]
		Given $\psi_{\bf e}(s_0{\bf s})$, a non-achromatic perceived color, its hue is the pure chromatic state $\varphi^*_{\bf e}({\bf s})$ defined by
		\begin{equation}\label{eq:hue}
			\varphi^*_{\bf e}({\bf s}):=\underset{\rho\in \mathcal{PS}(\mathcal H(2,\mathbb R))}{\arg\min}R(\rho||\rho_{\varphi_{\bf e}({\bf s})}).
		\end{equation}
	\end{definition}
	Notice that \textit{the hue of  $\psi_{\bf e}(s_0{\bf s})$ does not depend on $s_0$} because of the property $\varphi_{\bf e}(s_0{\bf s})=\varphi_{\bf e}({\bf s})$ for all $s_0\in [0,1]$.
	
	Of course, we must verify that the definition is well-posed, i.e. that the minimization problem defined by eq. \eqref{eq:hue} exists and it is unique. Thanks to the Klein inequality, the relative entropy is null if and only if $\rho=\varphi_{\bf e}({\bf s})$, so let us avoid this trivial case and also the achromatic condition (since achromatic colors lack of hue by definition) by supposing that $0< r_{\varphi_{\bf e}({\bf s})}<1$. 
	
	Let us notice that, thanks to the definition of relative entropy and to eq. \eqref{eq:Relentropy} we get:
	\begin{equation}
		\begin{split}
			R(\rho_{\bf s}||\rho_{\bf t}) & ={\rm Tr}(\rho_{\bf s}\log_2\rho_{\bf s}) - \Tr(\rho_{\bf s}\log_2\rho_{\bf t})\\ & =  -S(\rho_{\bf s})-\frac{1}{2}\log_2(1-r_{\bf t}^2)-\frac{r_{\bf s}\cos\vartheta_{{\bf s},{\bf t}}}{2}\log_2\left(\frac{1+r_{\bf t}}{1-r_{\bf t}}\right).
		\end{split}
	\end{equation}
	Now we must replace the generic density matrix $\rho_{\bf s}$ with one, $\rho$, associated to a pure state, so that $S(\rho_{\bf s})=0$ and $r_{\bf s}=1$, and $\bf t$ with $\varphi_{\bf e}({\bf s})$, thus obtaining:
	\begin{equation}\label{eq:100}
		\begin{split}
			R(\rho||\rho_{\varphi_{\bf e}({\bf s})}) & =-\frac{1}{2}\log_2(1-r_{\varphi_{\bf e}({\bf s})}^2)-\frac{\cos\vartheta_{\rho,\varphi_{\bf e}({\bf s})}}{2}\log_2\left(\frac{1+r_{\varphi_{\bf e}({\bf s})}}{1-r_{\varphi_{\bf e}({\bf s})}}\right).
		\end{split}
	\end{equation}
	Since $0< r_{\varphi_{\bf e}({\bf s})}<1$,  $-\frac{1}{2}\log_2(1-r_{\varphi_{\bf e}({\bf s})}^2)> 0$ and  $\log_2\left(\frac{1+r_{\varphi_{\bf e}({\bf s})}}{1-r_{\varphi_{\bf e}({\bf s})}}\right)> 0$. 
	Given that $R(\rho||\rho_{\varphi_{\bf e}({\bf s})}) >0$ and that $r_{\varphi_{\bf e}({\bf s})}$ is fixed, the computation of the $\arg\min$ in eq. \eqref{eq:hue} is equivalent to the maximization of $\cos \vartheta_{\rho,\varphi_{\bf e}({\bf s})}$, i.e. \textit{we can reformulate the definition of hue of $\psi_{\bf e}(s_0{\bf s})$ as follows}:
	\begin{equation}
		\varphi^*_{\bf e}({\bf s}):=\underset{\rho\in \mathcal{PS}(\mathcal H(2,\mathbb R))}{\arg\max}{\cos \vartheta_{\rho,\varphi_{\bf e}({\bf s})}}.
	\end{equation}
	Recalling that $\vartheta_{\rho,\varphi_{\bf e}({\bf s})}$ is the angle between the chromatic vectors ${\bf v}_{\rho}$ of the pure state defined by $\rho$ (so that $r_{\rho}=\|{\bf v}_\rho\|=1$) and ${\bf v}_{\varphi_{\bf e}({\bf s})}$, which is fixed, we get: 
	\beq 
	\cos\vartheta_{\rho,\varphi_{\bf e}({\bf s})}=\frac{{\bf v}_{\rho} \cdot {\bf v}_{\varphi_{\bf e}({\bf s})}}{r_{\varphi_{\bf e}({\bf s})}},
	\eeq  
	which is maximized when ${\bf v}_{\rho}$ is parallel to ${\bf v}_{\varphi_{\bf e}({\bf s})}$.
	
	Hence, given the perceived color $\psi_{\bf e}(s_0{\bf s})$, we consider the corresponding state $\varphi_{\bf e}({\bf s})=\psi_{\bf e}(s_0{\bf s})/\langle {\bf e} \rangle_{s_0\bf s}$ and we represent it via the density matrix
	\begin{equation}
		\rho_{\varphi_{\bf e}({\bf s})}= {1\over 2}\left(\begin{array}{cc}1+r_{\varphi_{\bf e}({\bf s})}\cos\vartheta_{\varphi_{\bf e}({\bf s})} & r_{\varphi_{\bf e}({\bf s})}\sin\vartheta_{\varphi_{\bf e}({\bf s})} \\r_{\varphi_{\bf e}({\bf s})}\sin\vartheta_{\varphi_{\bf e}({\bf s})} & 1-r_{\varphi_{\bf e}({\bf s})}\cos\vartheta_{\varphi_{\bf e}({\bf s})}\end{array}\right),
	\end{equation}
	then, its hue is the pure state $\varphi^*_{\bf e}({\bf s})$ identified by the density matrix
	\begin{equation}\rho_{\varphi^*_{\bf e}({\bf s})}={1\over 2}\left(\begin{array}{cc}1+\cos\vartheta_{\varphi_{\bf e}({\bf s})} & \sin\vartheta_{\varphi_{\bf e}({\bf s})} \\\sin\vartheta_{\varphi_{\bf e}({\bf s})} & 1-\cos\vartheta_{\varphi_{\bf e}({\bf s})}\end{array}\right).
	\end{equation}
	What just proven not only shows that our definition of hue is well-posed, but it is also in perfect agreement with the interpretation of pure states as hues already discussed in section \ref{subsec:chromentropies}.
	
	We  emphasize the fact that the two definitions of saturation and hue by means of relative quantum entropy are much more significant from the perception viewpoint than those involving \textit{ad hoc} coordinates of classical colorimetric spaces. The relative entropy between two states is a measure of their \textit{distinguishability}. This precisely means that \textit{the saturation of a perceived color is a measure of how it can be  distinguished from the achromatic state}. In the same way, \textit{the hue of a perceived color is the closest, from the distinguishability point of view, pure chromatic state to the given perceived color.} We also insist on the fact that the above computations make use of the Bloch parameters of the state space of the rebit which are not the coordinates of the color appearance models of the CIE.
	
	As a consequence, the novel definitions of perceptual attributes that we propose constitute not only a meaningful formalization of the CIE definitions given in section \ref{sec:vocabulary}, but they are also mathematically operative in the quantum-like framework discussed above. They provide a rigorous explanation of the intuitive representation that one may have of the perceived color solid. 
	
	To conclude, having defined a perceived color via a post-measurement generalize state, i.e.
	\beq
	\psi_{\bf e}(s_0 {\bf s}) = e_0 s_0(1+{\bf v_e} \cdot \bf v_{\bf s}) \, \varphi_{\bf e}({\bf s}) = \langle {\bf e} \rangle_{s_0 \bf s} \,  \varphi_{\bf e}({\bf s}), 
	\eeq 
	and its brightness as 
	\beq
	\mathcal{B}(\psi_{\bf e}(s_0 {\bf s}))=e_0 s_0(1+{\bf v_e} \cdot \bf v_{\bf s}) = \langle {\bf e} \rangle_{s_0 \bf s},
	\eeq 
	we can rewrite a perceived color in the more explicit form given by
	\beq 
	\psi_{\bf e}(s_0 {\bf s}) = \mathcal{B}(\psi_{\bf e}(s_0 {\bf s})) \,  \varphi_{\bf e}({\bf s}), 
	\eeq 
	which corresponds more closely with the usual way of describing a color with a one-dimensional achromatic component, $\mathcal{B}(\psi_{\bf e}(s_0 {\bf s}))$, and a two-dimensional chromatic one, which, in our case, is contained in  $\varphi_{\bf e}({\bf s})$.
	
	In the next section, we illustrate the potential of our new system of definitions on the specific example of the lightness constancy phenomenon.

	\section{Characterization of lightness constancy in the quantum-like framework}\label{sec:lightconstacr}
	In this section we analyze the important property of lightness constancy from the point of view of the quantum-like framework and we characterize it through a precise equation involving generalized states. 
	
	\subsection{The phenomenon of lightness constancy}
	\label{sec:lcphenom}
	In order to fix the ideas, we wish to quote the following description of lightness constancy offered by \cite{Krantz:site}, that we find well-suited for our purposes: `\textit{Lightness constancy refers to the observation that we continue to see an object in terms of the proportion of light it reflects rather than the total amount of light it reflects. That is, a gray object will be seen as gray across wide changes in illumination. A white object remains white in a dim room, while a black object remains black in a well-lit room. In this sense, lightness constancy serves a similar function as color constancy in that it allows us to see properties of objects as being the same under different conditions of lighting. Consider an object that reflects 25\% of the light that hits its surface. This object will be seen as a rather dark gray. If we leave it in a dim room that receives only 100 units of light, it will reflect 25\% units of light. However, if we place it in a room that is better lit, it will still reflect the same 25\%. If there are now 1,000 units of light, it will reflect 250 units of light. But we still see it as approximately the same gray, despite the fact that the object is reflecting much more light. Similarly, an object that reflects 75\% of ambient light will be seen as a light gray in the dim room, even though it reflects less total light than it does in the bright room. Thus, lightness constancy is the principle that we respond to the proportion of light reflected by an object rather than the total light reflected by an object}'.
	
	To visually illustrate the difference between lightness and brightness judgment, let us consider the scene depicted in Figure   \ref{fig:Fig1}: the horizontal stripes of the building on the left and on the right of the yellow entrance are built with the same material, thus they have the same reflectance, however, some parts are exposed to sunlight and some other are not, due to the shadow projected by the tree. 
	
	If we had to make a  \textit{brightness judgment}, we would describe them as brighter and dimmer, respectively. Instead, if we had to express a \textit{lightness judgment}, we would state that all of them are `white', implicitly meaning that the parts directly hit by sunlight and those covered by the tree shadow \textit{would appear identical if they were lit in the same way}. This is an instance of the lightness constancy property of the human visual system. 
	
	The same analysis can be repeated for the parts of the yellow entrance covered or not by the tree shadow. So, thanks to lightness constancy, an observer would exclude the possibility that the part of horizontal stripes or the entrance in shadow are painted with a darker shade of gray or yellow, respectively, but that the perceptual difference is merely due to a different intensity in the lighting condition.
	
	\begin{figure}[htbp]
		\begin{center}
			\includegraphics[scale=0.06]{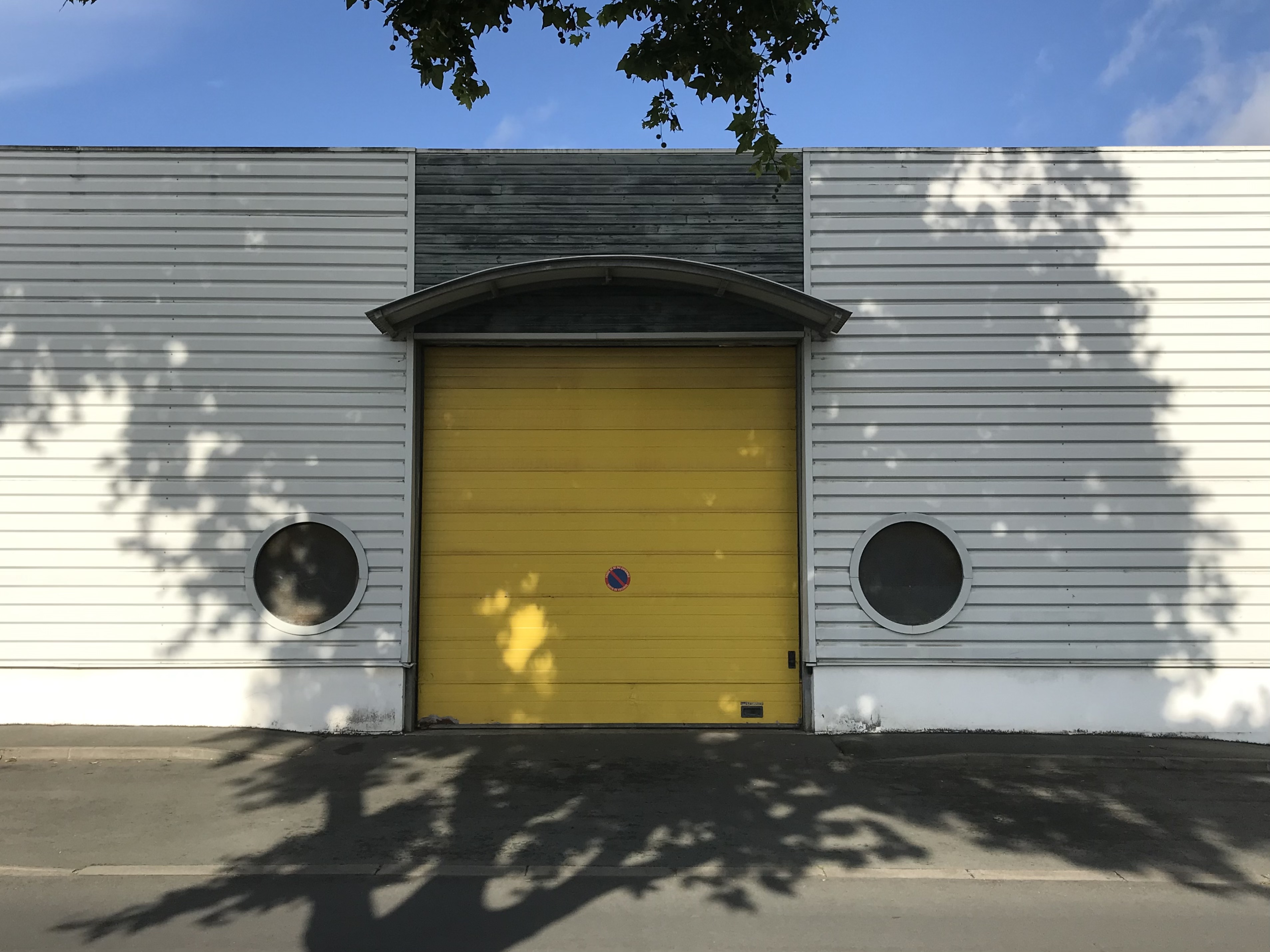}
			\caption{A visual scene used to illustrate the lightness constancy phenomenon.}
			\label{fig:Fig1}
		\end{center}
	\end{figure}
	
	
	The psycho-physiological reasons underling lightness constancy are still debated; we refer the reader to e.g. \cite{Ebner:07} for further information.
	
	\subsection{Lightness constancy derivation}
	\label{sec:lcderiv}	
	The two perceptual patches of interest are given by the two generalized states $p^1=p^1_0{\bf p}^1$ and $p^2=p^2_0{\bf p}^2$, where $p^1_0,p^2_0\in [0,1]$ and ${\bf p}^1,{\bf p}^2$ are two chromatic states. We first consider the case where the two (physical) effects corresponding to the two illuminants are achromatic, i.e. represented by $\biota^1=(\iota^1_0,{\bf 0})$ and $\biota^2=(\iota^2_0,{\bf 0})$. In classical colorimetry, this amounts to considering a D65 illuminant, see e.g. \cite{Wyszecky:82}.

	The two reflected lights of interest are thus given by the two generalized states
	\begin{equation}
		r^1=\psi_{\biota^1}(p^1_0{\bf p}^1)=\iota^1_0p^1_0{\bf p}^1,\ \ \ r^2=\psi_{\biota^2}(p^2_0{\bf p}^2)=\iota^2_0p^2_0{\bf p}^2.
	\end{equation}
	We consider now an observer $o$ associated to a (perceptual) achromatic effect $(o_0,{\bf 0})$. As explained before, this observer perceives the chromatic information of the two reflected lights `as they are', which means that there is no variation between the chromatic features of the reflected lights and the chromatic features of the perceived colors. More precisely, we have
	\begin{equation}
		\psi_{\bf o}(r^1)=o_0\iota^1_0p^1_0{\bf p}^1,\ \ \ \psi_{\bf o}(r^2)=o_0\iota^2_0p^2_0{\bf p}^2.
	\end{equation}
	
	It makes sense to consider the phenomenon of lightness constancy only when two perceived colors share the same chromatic information. In the present case, the chromatic matching equation \eqref{eq:chromatch1} leads trivially to ${\bf p}^1={\bf p}^2=:{\bf p}$. As a consequence,
	\begin{equation}
		r^1=\iota^1_0p^1_0{\bf p},\ \ \ r^2=\iota^2_0p^2_0{\bf p},
	\end{equation}
	and
	\begin{equation}
		\psi_{\bf o}(r^1)=o_0\iota^1_0p^1_0{\bf p},\ \ \ \psi_{\bf o}(r^2)=o_0\iota^2_0p^2_0{\bf p}.
	\end{equation}
	
	The two reflected lights $r^1$ and $r^2$ come from the two different illuminants $\iota^1$ and $\iota^2$, and the observer has to compensate the difference between these two illuminants in order to compare the initial perceptual patches $p^1$ and $p^2$, i.e. to compare the lightnesses $p^1_0$ and $p^2_0$. This means that the observer must find a way to recover the reflected lights as if they where lit by the same illuminant.
	
	Let the observer $o$ change his/her (perceptual) effect from $(o_0,{\bf 0})$ to $(o_0^1,{\bf 0})$ to define a new observer $o^1$ perceiving the reflected light $r^1$. In the same way, let the observer $o$ change his/her (perceptual) effect from $(o_0,{\bf 0})$ to $(o_0^2,{\bf 0})$ to define a new observer $o^2$ perceiving the reflected light $r^2$. We have 
	\begin{equation}
		\psi_{{\bf o}^1}(r^1)=o^1_0\iota^1_0p^1_0{\bf p},\ \ \ \psi_{{\bf o}^2}(r^2)=o^2_0\iota^2_0p^2_0{\bf p}.
	\end{equation}
	These two perceived colors are those obtained from a measurement of {\em only one observer associated to an achromatic effect} from the reflected lights produced by the two perceptual patches $p^1$ and $p^2$ {\em lit with the same achromatic illuminant} if and only if $o^1_0\iota^1_0=o^2_0\iota^2_0$. If, for instance, $o^1=(\iota^2_0,{\bf 0})$ and $o^2=(\iota^1_0,{\bf 0})$, then
	\begin{equation}
		\psi_{{\bf o}^1}(r^1)=\iota^2_0\iota^1_0p^1_0{\bf p},\ \ \ \psi_{{\bf o}^2}(r^2)=\iota^1_0\iota^2_0p^2_0{\bf p}
	\end{equation}
	are the perceived colors measured by the observer $o=(1,{\bf 0})$ from the reflected lights $\widetilde r^1=\iota_0^1\iota_0^2p_0^1{\bf p}$ and $\widetilde r^2=\iota_0^1\iota_0^2p_0^1{\bf p}$ obtained by illuminating the perceptual patches $p^1$ and $p^2$ with the same achromatic illuminant $(\iota_0^1\iota_0^2,{\bf 0})$. Using the equation $o^1_0\iota^1_0=o^2_0\iota^2_0$, it appears clearly that the two perceived colors $\psi_{{\bf o}^1}(r^1)$ and $\psi_{{\bf o}^2}(r^2)$ are equal if and only if the two lightnesses $p_0^1$ and  $p_0^2$ are equal.
	
	The above analysis of lightness constancy requires two measurements performed by the observers $o^1$ and $o^2$, with the condition $o_0^1/o_0^2=\iota_0^2/\iota_0^1$, in order to make some comparison. This means that the lightness constancy phenomenon requires that the observer is able to evaluate the ratio $\iota_0^2/\iota_0^1$ between the two magnitudes of the illuminants $\iota^2$ and $\iota^1$.
	
	One can check that the proposed derivation also applies when the two illuminants are no more achromatic but still share the same chromatic features expressed by their effect vector.

	\section{Conclusions}\label{sec:conclusions}
	
	Our will to formalize the definition of chromatic attributes was guided by the following inspiring words of J. Clerk Maxwell: `\textit{The first process, therefore, in the effectual study of the sciences, must be one of simplification and reduction of the results of previous investigations to a form in which the mind can grasp them}', see e.g. \cite{Jauch:68}. 
	
	In fact, paraphrasing this quotation, it was our need to grasp the precise mathematical meaning of perceptual color attributes that led us to study  `previous investigations' on this topic and adapt them to the quantum-like color perception framework.
	
	Thanks to the great versatility of quantum effects in relation with color measurements and to the fundamental concept of generalized state, we were able to propose rigorous definitions for the perceptual attributes of color sensations generated by both light sources and by non-emitting surfaces lit by an illuminant. 
	
	The novel definitions of perceptual attributes provide a rigorous explanation of the intuitive representation that one may have of the perceived color solid, they are mathematically operative in the quantum-like framework and they constitute a meaningful formalization of the CIE definitions. This last consideration, in particular, underlines the potential of our results in color imaging, where it is well-known that the processing of intrinsically perceptual attributes, as e.g. hue and saturation, with the tools offered by the classical color spaces requires extreme caution to avoid color artifacts. This often results in sub-optimal algorithms that could clearly benefit from the use of perceptually-coherent colorimetric quantities.
	
	
	
	An important future step consists in dealing with related colors. For that, the quantum-like framework must be non-trivially modified to mathematically model color induction phenomena, a problem that we are currently studying. 
	
	We are also interested in a quantitative description of colorimetric effects, as e.g. the Bezold-Brücke and Helmholtz-Kohlrausch ones \cite{Fairchild:13}, whose deep understanding may be beneficial in particular in HDR imaging, where these effects are magnified by the large range of intensity provided by HDR screens, or in the still open problem of tone mapping. Finally, we are interested in studying novel perceptual chromatic metrics starting from the  Hilbert-Klein distance.
	
	The results that we have obtained seem to confirm an auto-consistent theory without incoherences able to make quantitative predictions which can be tested with experiments. This is the best that we can hope to get without the much needed help of empirical data to support or disprove our model, or to guide us toward adjustments and improvements. In this sense, we hope that our work can be taken as a source of inspiration for experimentalists to perform novel perceptual tests.

	\bibliographystyle{plain} 
	\bibliography{bibliography}

\end{document}